\documentclass[prl,amsmath,amssymb,groupedaddress,superscriptaddress,floatfix,twocolumn,showkeys,amsart]{revtex4-1} 

\usepackage{array,amsmath,amsfonts,amssymb,dsfont,tabularx,multirow,blkarray,bigstrut}

\usepackage{hhline}
\usepackage[T1]{fontenc} \usepackage{ae} \usepackage{color}
\usepackage{bbold}
\usepackage{enumitem}
\usepackage{lipsum}

\newcolumntype{C}[1]{>{\centering\arraybackslash$}p{#1}<{$}}

\usepackage{graphicx}
\usepackage{soul} 
\usepackage{balance}

\usepackage{amssymb}
\usepackage{physics}
\usepackage{pifont}

\newcommand{\appropto}{\mathrel{\vcenter{
  \offinterlineskip\halign{\hfil$##$\cr
    \propto\cr\noalign{\kern2pt}\sim\cr\noalign{\kern-2pt}}}}}

\usepackage{makecell,tabularx}

\renewcommand{\i}{{\mathrm i}} \def\1{\mathchoice{\rm 1\mskip-4.2mu l}{\rm 1\mskip-4.2mu l}{\rm
1\mskip-4.6mu l}{\rm 1\mskip-5.2mu l}}

 \newcommand{\id}

\usepackage{gensymb} 

\begin{document}

\title{Coherence of Rabi oscillations with spin exchange}

\author{C.~Kiehl}\email{christopher.kiehl@colorado.edu}
\affiliation{JILA, National Institute of Standards and Technology and University of Colorado, and
Department of Physics, University of Colorado, Boulder, Colorado 80309, USA}
\author{D.~Wagner} 
\affiliation{JILA, National Institute of Standards and Technology and University of Colorado, and
Department of Physics, University of Colorado, Boulder, Colorado 80309, USA}
\author{T.-W.~Hsu} 
\affiliation{JILA, National Institute of Standards and Technology and University of Colorado, and
Department of Physics, University of Colorado, Boulder, Colorado 80309, USA}
\author{S.~Knappe} 
\affiliation{Mechanical Engineering, University of Colorado, Boulder, Colorado 80309, USA}
\affiliation{FieldLine Inc., Boulder CO 80301, USA}
\author{C.~A.~Regal} 
\affiliation{JILA, National Institute of Standards and Technology and University of Colorado, and
Department of Physics, University of Colorado, Boulder, Colorado 80309, USA}
\author{T.~Thiele} 
\altaffiliation[Current address: ]{Zurich Instuments AG, Technoparkstrasse 1, 8005 Zurich, Switzerland}
\affiliation{JILA, National Institute of Standards and Technology and University of Colorado, and
Department of Physics, University of Colorado, Boulder, Colorado 80309, USA}

\begin{abstract}
Rabi measurements in atomic vapor cells are of current interest in a range of microwave imaging and sensing experiments, but are increasingly in a parameter space outside of theoretical studies of coherence defined by spin-exchange collisions. Here, we study the coherence of Rabi oscillations in vapor cells by employing continuous non-destructive readout of the hyperfine manifold of $^{87}$Rb using Faraday rotation. We develop a full model for spin-exchange (SE) coherence for hyperfine transitions that takes into account a non-static population distribution. In this regime, Rabi oscillations exhibit nontrivial time-domain signals that allow verification of vapor-cell parameters. We find excellent agreement between theory and experiment, which will aid in benchmarking sensitivities of Rabi measurement applications.

\end{abstract}
    
\maketitle
\raggedbottom
\textit{Introduction.}---For sensors based on hot atomic vapors, spin-exchange (SE) collisions are the dominant limitation of sensitivity~\cite{vanier1989quantum,happer1972optical}. These collisions originate from the acquired phase shift between singlet and triplet interaction potentials of the electrons of colliding alkali atoms. The exchange interaction conserves the total spin of the colliding atoms, but causes random transitions between the hyperfine ground states that have recently been leveraged in hot atomic vapor cells for generating many-body entanglement~\cite{kong2020measurement} and modeling phase transition dynamics~\cite{horowicz2021critical}. Decoherence effects of SE collisions within a Zeeman manifold have been studied in the context of optically-pumped magnetometers (OPMs), and notably even found to disappear in the spin-exchange-relaxation-free (SERF) regime at low magnetic fields~\cite{Budker2007optical,happer1973spin,savukov2005effects,happer1973spin}. Further, the consequences of spin-exchange collisions on the coherence between two hyperfine ground manifolds have also been well-studied in the context of masers~\cite{vanier1968relaxation,vanier1974relaxation} and atomic clocks~\cite{jau2004intense,vanier1989quantum}. In these studies, SE decay rates were modeled based on stationary atomic populations such as a spin-temperature (ST) distribution, which is valid when continuous optical pumping and a weak driving field prevent Rabi oscillations, and SE collisions dominate over other collision and scattering rates.
However, this approximation is invalid in the case of a strong driving field, such as a near-resonant microwave field, that causes significant population transfer. As a result, the coherence of Rabi oscillations is expected to deviate from the assumptions of weak driving and exhibit nontrivial detuning and spin-polarization dependence. 

This picture needs completion, as in the last few years, Rabi oscillations driven within atomic ensembles have been proposed for sensing and imaging microwave fields for applications in characterizing microwave circuits~\cite{horsley2015widefield,horsley2016frequency-tunable,bohi2012simple}, self-calibrated vector magnetometry~\cite{thiele2018self}, and in medical applications such as cancer detection~\cite{fear2002enhancing,nikolova2011microwave,chandra2015opportunities}. While microfabricated vapor cells are ideal sensor platforms for these applications due to their compact size and high atomic densities, decoherence is typically dominated by SE collisions when wall collisions are minimized by buffer gas~\cite{wittke1956redetermination}. To unlock new sensitivity regimes for these applications, a complete understanding of the SE coherence is necessary.

\begin{figure}[tbh]\centering
\includegraphics[width=0.45\textwidth]{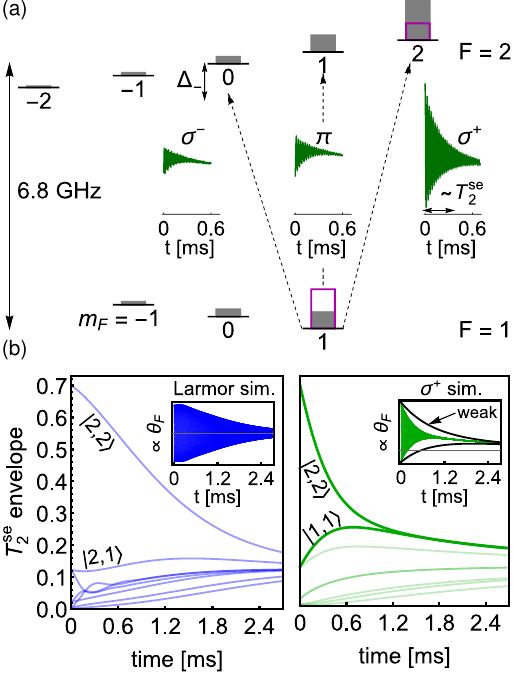}
\caption{ (a) Energy-level diagram for $^{87}$Rb showing the relevant microwave transitions with Rabi oscillations measured in our apparatus (green insets). A microwave sweep performs adiabatic rapid passage (ARP) to switch the $\sigma^+$ populations prior to driving the $\sigma^-$ and $\pi$ transitions (magenta). (b) The envelopes of simulated atomic population dynamics due purely to SE collisions during  Larmor precession (left) which decays slower than during a $\sigma^+$ Rabi oscillation (right). Comparison of the envelopes for each state population to the full-simulated Faraday rotation signal (insets) shows nuanced population dynamics that are easier to individually observe with Rabi oscillations that couple a pair of states. The initial populations and SE rate here are set by a ST distribution with polarization p $=0.7$ and vapor temperature $\mathcal{T}_v=110^{\circ}$C}
\label{fig:Figure1main}
\end{figure}

In this Letter, we explore the coherence of Rabi oscillations in a heated vapor cell driven on $\sigma^-,\pi$, and $\sigma^+$ hyperfine transitions of $^{87}$Rb [Fig.~\ref{fig:Figure1main}(a)] that do not adhere to the weak-driving approximation. By employing continuous quantum non-demolition measurement readout based on Faraday rotation we are able to study nontrivial time-dependent behavior of the atomic population. Using a full theoretical analysis of SE that accounts for the time-dependent SE dephasing rate caused by population dynamics during Rabi oscillations, we observe excellent agreement between the model and the measured coherence in the continuous Faraday signal.

With this full understanding in hand, we illustrate that the Rabi lineshape is connected to many vapor parameters and show that driving multiple transitions can pinpoint useful information. Specifically, we extract from SE coherence consistent values for the vapor temperature, buffer gas pressure, and the atomic state prepared by optical pumping by using the fact that the Rabi envelope reflects the initial atomic populations, as well as their subsequent SE redistribution predicted from our model. While similar population dynamics occurs with OPMs that sense the free Lamor precession, such SE effects are more apparent and distinguishable with Rabi oscillations that probe discrete states [Fig.~\ref{fig:Figure1main}(b)]. 

\textit{Theoretical model.}---We first describe our theoretical model for driving Rabi oscillations in an alkali vapor cell. The relevant atomic dynamics occurs in the two hyperfine ground manifolds $F=I\pm 1/2$. Hence we describe the state of the atomic ensemble by a $(4I+2)\times(4I+2)$ density matrix in the $\ket{F,m_F}$ basis. The time evolution of the atomic ensemble including multiple sources of collisional decoherence is given by~\cite{budker2013optical}
\begin{align}
\label{eq:timeEvo}
\begin{split}
    \frac{d\rho}{dt}=&\frac{[H(t),\rho]}{i \hbar}+\Gamma_{\text{se}}\big(\phi(1+4\langle \textbf{S} \rangle\cdot \textbf{S})-\rho\big)+\\&\Gamma_{\text{sd}}\big(\phi-\rho\big)-\frac{\eta_I^2 [I]}{8}\Gamma_{\text{C}} \rho^{(\text{m})}+  D\nabla^2\rho
\end{split}
\end{align}
where $\phi=\rho/4+\textbf{S}\cdot \rho\textbf{S}$ is known as the nuclear part of the density matrix for which Tr$[\phi \textbf{S}]=0$ and Tr$[\phi]=1$ hold~\cite{baranga1998polarization}. The Hamiltonian
\begin{equation}
\label{eq:Hamiltonian}
    H(t)=A_{\text{hfs}}\textbf{I}\cdot \textbf{S}+\mu_B(g_s \textbf{S}+g_I \textbf{I})\cdot(B^{\text{dc}}\hat{z}+\textbf{$\mathcal{B}$}^{\mu w}(t))
\end{equation}
describes the hyperfine structure, Zeeman shift from an applied DC magnetic field oriented along $\hat{z}$, and the atom-microwave coupling. Here,
\begin{equation}
    \textbf{$\mathcal{B}$}^{\mu w}(t)=\sum_{i=x,y,z}\mathcal{B}_i\cos(\omega_{\mu w}t+\phi_i)\hat{e}_i
\end{equation}
describes the magnetic part of the microwave field~\cite{kopsell2017measuring} that drives Rabi rates $\Omega_{j}=\mu_j \mathcal{B}_j/\hbar$ defined in the spherical basis $j\in\{\pm,\pi \}$ with transition dipole moments $\mu_j$. To eliminate the large energy scale caused by $A_{\text{hfs}}$, as well as high-frequency counter-rotating terms in Eq.~\eqref{eq:Hamiltonian}, we assume the rotating wave approximation to reduce the computation time required to solve Eq.~\eqref{eq:timeEvo}. Full details on the Hamiltonian and the RWA are discussed in the Supplementary Material~\cite{suppMat}.

\begin{table}[t]
\caption{\label{tab:collisionRates}%
Cross-sections, diffusion constant, and the corresponding collision rates $\Gamma$ for the different collisional processes for a vapor cell with volume ($3\times3\times2$ mm$^3$), temperature $\mathcal{T}_v=108^{\circ}$C, buffer gas pressure $\text{P}_{\text{N}_2}=180$ Torr, and diffusion constant $D_0= 0.221$ cm$^2/s$ for Rb-N$_2$ buffer gas collisions~\cite{pouliot2021accurate}. We use the known $D_0 \propto \mathcal{T}_v^{3/2}$ dependence to scale $D_0$ to our cell temperature. The wall collision rate reported is for the fundamental diffusion mode~\cite{franzen1959spin}.}
\begin{ruledtabular}
\begin{tabular}{ccc}
\textrm{Collision} &
\textrm{$\sigma$ $[10^{-18} \text{ m}^2]$}&
\textrm{$\Gamma=n\sigma v^r$ [Hz]}\\
\colrule
(Rb-Rb)$_{\text{se}}$  & 1.9~\cite{micalizio2006spin} & 6.6$\times 10^3$\\
(Rb-Rb)$_{\text{sd}}$ & $1.77\times 10^{-4}$~\cite{baranga1998polarization} & 6.1\\
($\text{N}_2$-Rb)$_{\text{sd}}$ & $1.44\times 10^{-8}$~\cite{chupp1989laser} & 40\\
($\text{N}_2$-Rb)$_{\text{carver}}$ & $\Gamma_{\text{C}}/[\text{N}_2]=394 \text{ amg}^{-1} \text{s}^{-1}$~\cite{walter2002magnetic} & $\Gamma_{\text{C}}=67$\\
wall & $D_0P_0=0.017 \text{ m}^2\text{Torr}$~\cite{pouliot2021accurate} &  $\Gamma_{\text{D}}=4.4\times10^2$\\

\end{tabular}
\end{ruledtabular}
\end{table}

Coherence effects due to SE and S-damping (SD) collisions are modeled by the  second and third terms of Eq.~\eqref{eq:timeEvo}, respectively. Collision rates $\Gamma_{\text{se(sd)}} \equiv n_{a}\sigma_{\text{se(sd)}}v^{r}$ are characterized by cross sections $\sigma_{\text{se(sd)}}$, the mean relative velocity of the colliding pair $v^r$, and $n_{a}$ $(n_{\text{buff}})$ the atomic density for alkali (buffer gas) collisions~\cite{allred2002high-sensitivity}. The fourth term models pure dephasing of the microwave transitions due to buffer gas collisions where $\rho^{(m)}$ represents the density matrix with only off-diagonal terms of the coherences between the upper and lower hyperfine manifolds, $\Gamma_{\text{C}}$ is the Carver rate, and $\eta_{I}=\mu_I/2I\mu_{\text{N}}$ is the isotope coefficient defined by the nuclear magnetic moment $\mu_I$ and nuclear magneton $\mu_{\text{N}}$~\cite{jau2005new,walter2002magnetic}.  The fifth term of Eq.~\eqref{eq:timeEvo} models diffusion into the cell wall where  alkali spins are completely randomized. Here $D=D_0 \text{P}_0/\text{P}_{\text{buff}}$ is the diffusion constant attenuated by the buffer gas pressure where $\text{P}_0=1$ atm. We approximate the effect of wall collisions by replacing $D\nabla^2\rho\rightarrow  \Gamma_{\text{D}}(\rho^e-\rho)$, where $\rho^e$ is the diagonalized density matrix with all populations $\rho^e_{ii}$ equal and $\Gamma_{\text{D}}=D\pi^2/(l_x^2+l_y^2+l_z^2)$ is the fundamental decay mode defined by the vapor cell dimensions $l_x$, $l_y$, and $l_z$~\cite{franzen1959spin}. In practice, higher-order diffusion modes that cause multiple decay rates also need to be considered. For context, Table~\ref{tab:collisionRates} contains the collision rates, cross-sections, and diffusion constant assuming the vapor cell parameters and the Rb-N$_2$ alkali-buffer gas mixture used in our experiment.

First, we study the dephasing rate $\gamma\equiv 1/T_2$ of Rabi oscillations  with the atomic ensemble initialized in the steady-state solution when SE collisions dominate over other types of collisions and scattering rates. This solution, known as a spin-temperature distribution, is defined by the atomic populations as $\rho_{ii}(0)=\rho_{ii}^{(\text{p})}\propto e^{\beta(\text{p})m_{i}}$, where $\beta(\text{p})=\text{ln}\big[\frac{1+\text{p}}{1-\text{p}}\big]$ and $\text{p}\in[0,1]$ is the electron spin polarization. This situation is a close approximation for many vapor cell experiments that optically pump with high buffer gas pressure~\cite{jau2004intense,baranga1998polarization}. We specifically study the $\sigma^{\pm}$ and $\pi$ transitions in Fig.~\ref{fig:Figure1main}. For the $\pi$ and $\sigma^-$ Rabi oscillations, we set the initial atomic state away from a ST distribution in our simulation by switching the atomic populations of the $\sigma^+$ transition [Fig.~\ref{fig:Figure1main}(a) magenta boxes]. This increases the atomic population difference in the $\pi$ and $\sigma^-$ transitions and enables larger signals.

\begin{figure}[tbh]
\includegraphics[width=0.45\textwidth]{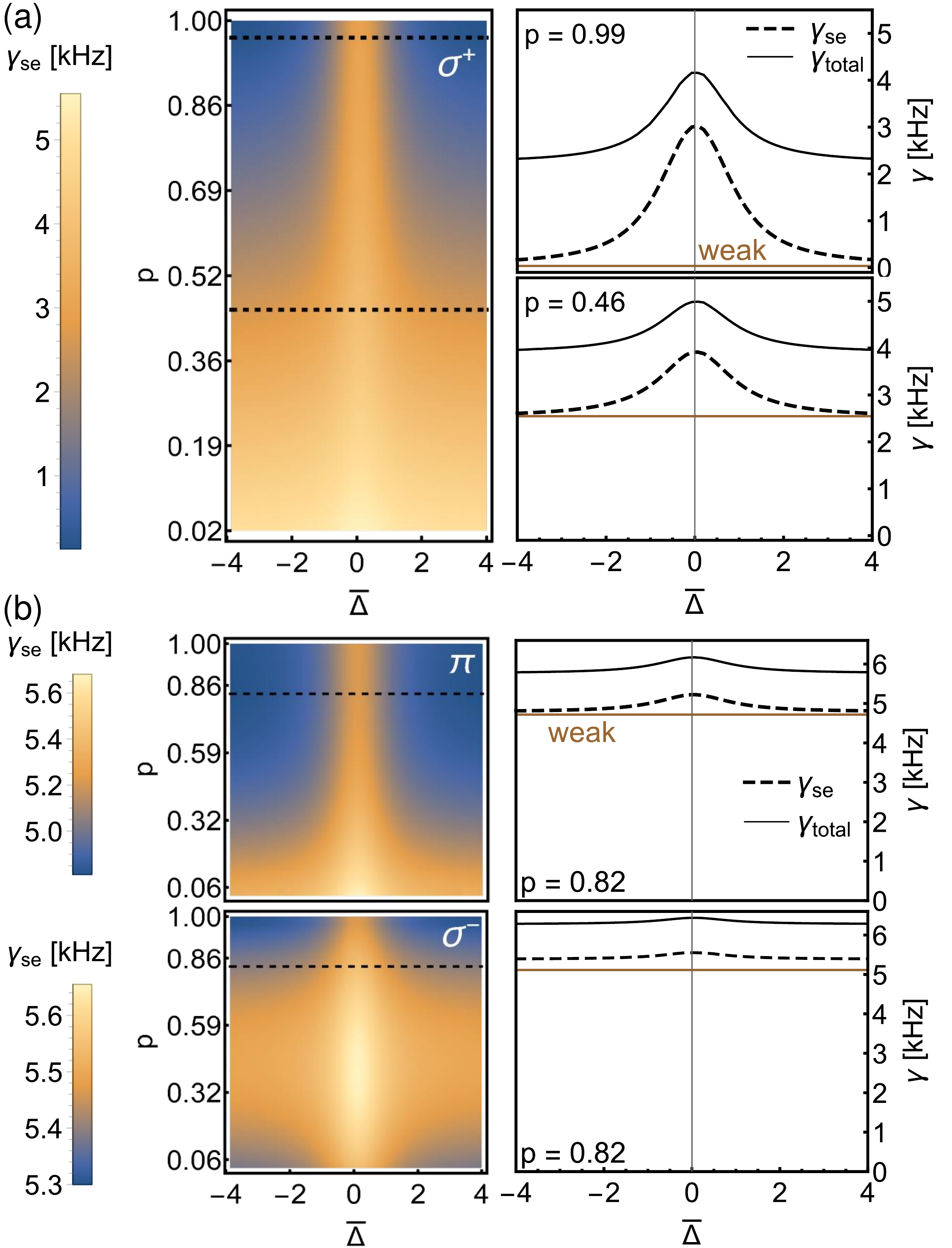}
\caption{Numerical simulation of the dependence of Rabi coherence with electron spin polarization (p) and normalized microwave detuning ($\overline{\Delta}$). The left density plots show the SE dephasing rate $\gamma_{\text{se}}$ with black dashed lines indicating p-cuts for the corresponding plots on the right. The right plots simultaneously show the $\gamma_{\text{se}}^{\text{wk}}$ (brown), the exact SE dephasing rate (black dashed), and the total decay rate $\gamma_{\text{total}}$ including SD, pure dephasing from buffer gas collisions, and wall collisions (black solid). For these plots we only include the fundamental diffusion mode to model wall collisions. The cell parameters used in these plots are $\mathcal{T}_{\text{v}}=110^{\circ}$C and $\text{P}_{\text{N}_2}=100$ Torr. (a) $\sigma^+$ transition with two p-cuts (b) $\pi$, $\sigma^-$ transitions each with one p-cut.}
\label{fig:Figure2main}
\end{figure}

For spin-exchange collisions, the dephasing rate $\gamma_{se}\propto \Gamma_{\text{se}}$ is proportional to the collision rate appearing in Eq.~\eqref{eq:timeEvo} and further depends on the atomic populations $\rho_{ii}(t)$ due to the different projections of the $\ket{F,m_F}$ basis on the electron spin. Hence, using Eq.~\eqref{eq:timeEvo}, we investigate $\gamma_{\text{se}}$ as a function of the electron spin-polarization $\text{p}$ and the normalized microwave detuning $\overline{\Delta}_j=\Delta_j/\Omega_j$ where $j=\pm,\pi$ as the primary variables affecting atomic population dynamics [Fig.~\ref{fig:Figure2main}]. For now we assume vapor parameters $\mathcal{T}_v=110\text{ }\degree$C and $\text{P}_{\text{N}_2}=100$ Torr and consider Rabi oscillations between the hyperfine ground states of $^{87}$Rb.

By neglecting the last three terms in Eq.~\eqref{eq:timeEvo}, we can compare the dephasing rate due to SE alone (black dashed) to the case with all collisional processes (black solid) as shown in Fig.~\ref{fig:Figure2main}. We also plot the predictions of the weak-driving approximation $\gamma_{\text{se}}^{\text{wk}}$ (brown) to show the large discrepancy between previous studies of SE linewidths with the full-numerical solution shown in this work in the regime of strong coherent driving. Such coherent dynamics and atomic population redistribution from SE collisions leads to an increased dephasing rate $\gamma_{se}$ near $\overline{\Delta}_j=0$ that is symmetric with $\overline{\Delta}_j$ for the transitions and initial states considered here [Fig.~\ref{fig:Figure2main}]. This effect is heightened for the $\sigma^+$ transition where no hyperfine-changing collisions occur for atoms purely in the initial \lq stretched\rq \text{ }state $\ket{2,2}$, when all electron spins are aligned, but are coherently coupled into $\ket{1,1}$ with projections onto states with oppositely aligned electron spins. If the initial atomic populations are not a ST distribution, then the hyperfine coherence will have multiple decay rates because the atomic populations are time-dependent even in the far-detuned limit. This explains the small descrepancy between the weak-driving approximation and the far-detuned Rabi driving in the case of the $\pi$ and $\sigma^-$ transitions due to the $\sigma^+$ population switch [Fig.~\ref{fig:Figure2main}(b)].

\begin{figure}[tbh]\centering
\includegraphics[width=0.45\textwidth]{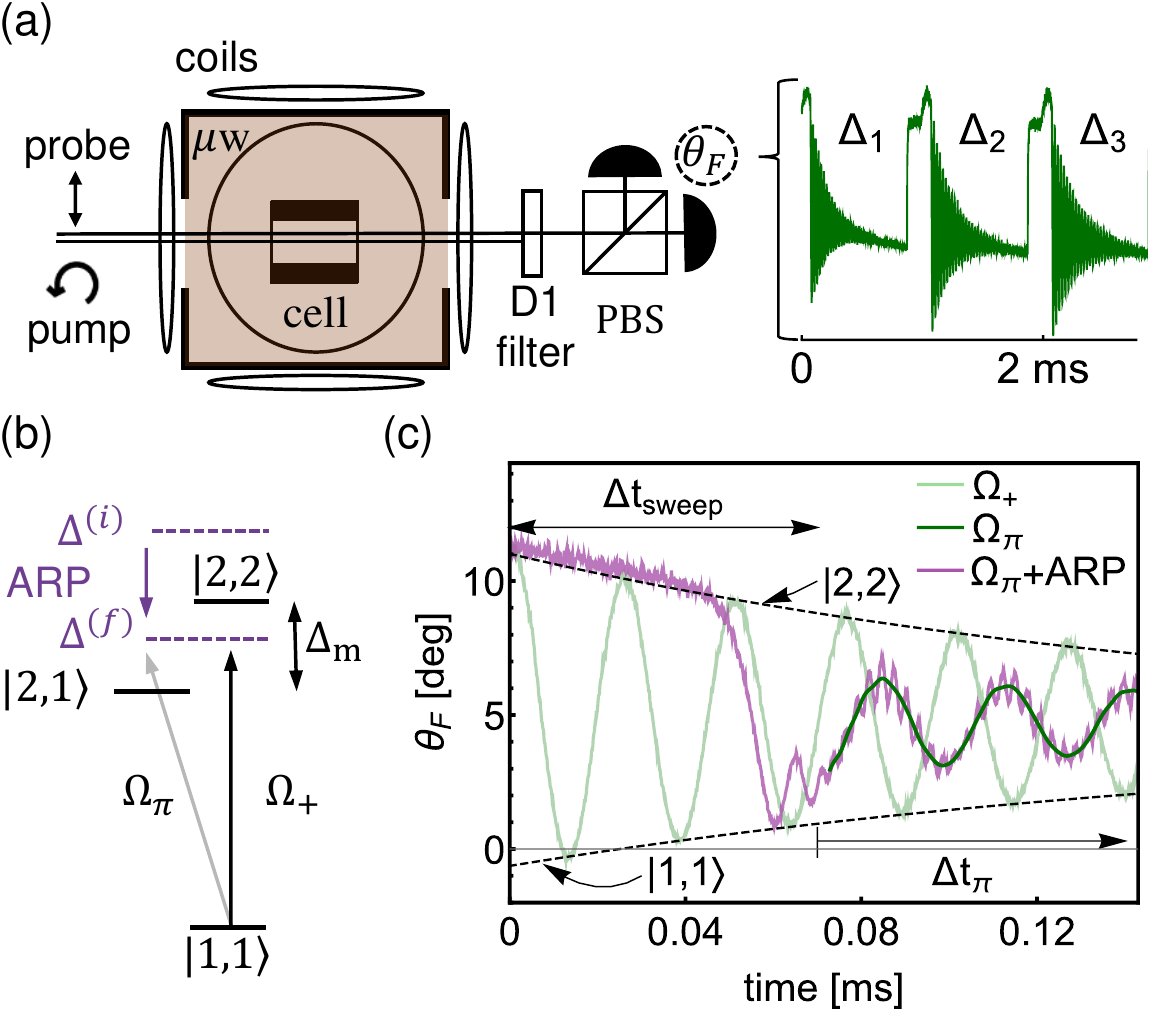}
\caption{(a) Apparatus for driving Rabi oscillations. We drive and measure Rabi oscillations between hyperfine states $\ket{1,1}$ to $\ket{2,0}$, $\ket{2,1}$, and $\ket{2,2}$ at multiple detunings in real-time using Faraday rotation readout of the macroscopic atomic spin state. (b) Energy level diagram for adiabatic rapid passage (ARP). (c) Using ARP to maximize the $\ket{1,1}$ population and enhance the $\Omega_{\pi}$ signal. Overlaying with $\Omega_{+}$ (light green) denotes $\theta_F$ values for $\ket{2,2}$ and $\ket{1,1}$ occupation, where we observe ARP transfer (purple) after the microwave frequency sweep. Extracting the pure $\pi$ oscillation (dark green) requires filtering off-resonance driving of adjacent transitions.}
\label{fig:Figure3main}
\end{figure}

\textit{Experimental Methods.}---To validate this theoretical description, we experimentally drive Rabi oscillations on the $\sigma^{\pm}$ and $\pi$ microwave transitions [Fig.~\ref{fig:Figure1main}(a)] near $\overline{\Delta}_j=0$, where the assumption of weak-driving does not hold, and further, where the theoretical model predicts larger SE dephasing rates than the weak-driving approximation. Our apparatus [Fig.~\ref{fig:Figure3main}(a)] consists of a square-shaped ($4.8\times4.8\times 2$ cm$^3$) microwave cavity with the degenerate modes ($n_x,n_y,n_z$)=$(2,1,0)$ and $(1,2,0)$ tuned near the $^{87}$Rb hyperfine ground state resonance ($6.834$ GHz), a microfabricated vapor cell ($3\times3\times2$ mm$^3$) filled with $^{87}$Rb and $\text{P}_{\text{N}_2}\approx$ 180 Torr buffer gas, and a DC coil system consisting of three pairs of near-orthogonal coils to create a static magnetic field ($\approx 50$~$\mu$T) to lift the degeneracy between adjacent magnetic sublevels by $\Delta_m \approx 350$ kHz [Fig.~\ref{fig:Figure3main}(b)]. While the use of a cavity is not crucial for this investigation, we plan to exploit the cavity modes and the Rabi spectroscopy presented here for future control of microwave polarization for vector magnetometry~\cite{thiele2018self}. Electrical heat tape heats the microwave cavity, and by design, also the cell to near 108$^{\circ}$C. To avoid stray magnetic fields, we turn off the heat tape 80 ms before each measurement. Due to the polarization structure of the microwave field, we tilt the magnetic field from the pump-beam axis by 25$^\circ$ such that a non-zero microwave field component drives all microwave hyperfine transitions with respect to $\ket{1,1}$ [Fig.~\ref{fig:Figure1main}(a)].

To study the coherent population transfer, we infer the atomic state by measuring the Faraday rotation angle ($\theta_F$), of the probe beam $\approx80$ GHz detuned from the D$_2$ line (780 nm) to obtain fast continuous measurements while minimizing decoherence from scattered light. While OPMs also typically use Faraday rotation to sense Larmor precession~\cite{allred2002high-sensitivity,shah2010high,li2011optical}, all previous measurements of Rabi oscillations in vapor cells, to our knowledge, have used absorptive imaging to sense spin dynamics~\cite{affolderbach2015imaging,bohi2012simple,horsley2015widefield,liu2018survival}. Our readout enables measurements of consecutive Rabi oscillations for various microwave detunings ($\Delta_{1}$, $\Delta_{2}, \dots$) and transitions in real time [Fig.~\ref{fig:Figure3main}(a)]. 

In each measurement, we first prepare a spin-polarized atomic ensemble by continuously optically pumping with circularly-polarized light that is near-resonant with the D$_1$ line (795 nm) for 100~$\mu$s. Immediately after, we adiabatically turn off the pumping light over the next 100~$\mu$s to align the atomic spins with the static magnetic field and avoid Larmor precession. After preparation of the macroscopic atomic spin in the $\ket{2,2}$ state, we prepare the ensemble in the $\ket{1,1}$ state using adiabatic rapid passage (ARP) [Fig.~\ref{fig:Figure3main}(b-c)]. For this we switch on a strong microwave drive at detuning $\Delta^{(i)}=490$ kHz above the $\sigma^+$ resonance and then linearly chirp it to $\Delta^{(f)}=-150$ kHz below the $\sigma^+$ resonance within $\Delta t_{\text{sweep}}=70$ $\mu$s. Then, we continue to study the dynamics of Rabi oscillations near-resonance by tuning to $\overline{\Delta}_j\approx 0$.  In the acquired data, we filter out high-frequency oscillations due to off-resonant coupling to adjacent microwave transitions [Fig.~\ref{fig:Figure3main}(c)]. Note, to study the $\sigma^+$ transition that couples $\ket{1,1}$ to $\ket{2,2}$, we do not perform an ARP.

\begin{figure*}[tbh]\centering
\includegraphics[width=\textwidth]{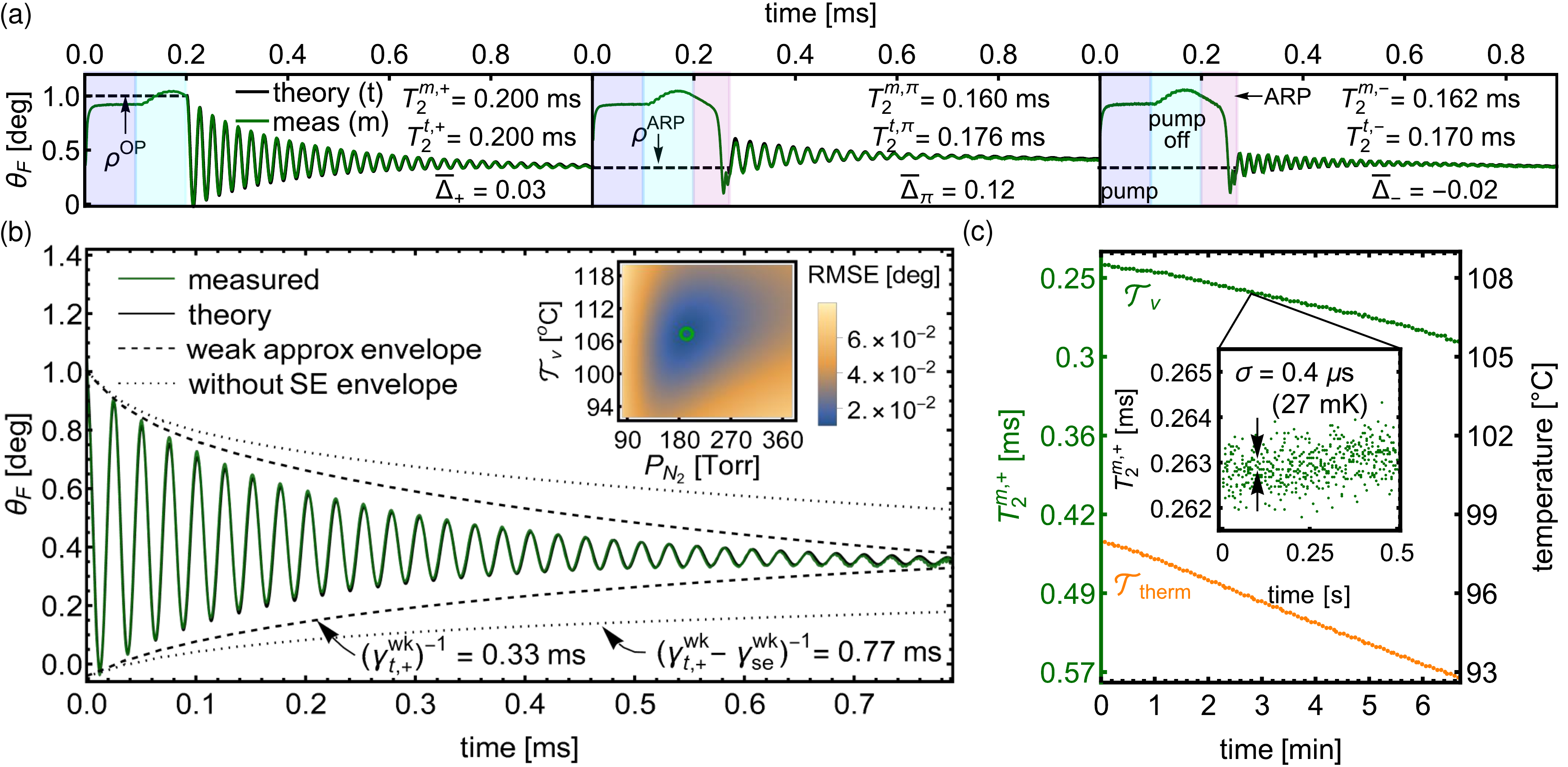}
\caption{ (a) A mirror comparison of the simulated (black) and measured (green) $\sigma^+$, $\pi$, and $\sigma^-$ Rabi oscillations. (b) A plot of the $\sigma^+$ Rabi oscillation alongside other relevant decay rates. Importantly, the weak-driving approximation predicts much higher coherence compared to our measurements. The inset shows the RMSE between the simulated and measured Rabi oscillations for different vapor temperatures $\mathcal{T}_v$ and buffer gas pressures $\text{P}_{\text{N}_2}$ at a fixed spin polarization p $=0.79$. By finding these optimal values, our model estimates these parameters for our cell. (c) Intra-cell thermometry with SE coherence measurements. The intra-cell temperature $\mathcal{T}_v$ is inferred from measured dephasing times $T^{m,+}_2$ using the calibrated model. The inset displays $T^{m,+}_2$ fluctuations over a 0.5 second period.}
\label{fig:Figure4main}
\end{figure*}

\textit{Verification of theoretical model}--- To assess the accuracy of the theoretical model, we mimic the experimental sequence and parameters when solving Eq.~\eqref{eq:timeEvo} to compare the theoretical and measured  dephasing rates and Rabi envelopes for the $\sigma^{\pm}$ and $\pi$ transitions as shown in Fig.~\ref{fig:Figure4main}(a-b). During this process we fit free parameters in our model that represent the vapor temperature $\mathcal{T}_v$, buffer gas pressure $\text{P}_{\text{N}_2}$, and the electron spin polarization $\text{p}$ produced by optical pumping, whose extracted values are consistent with independent knowledge of the apparatus. For modeling state preparation, we make the approximation that there are no initial atomic coherences and use the Rabi measurements to fit the atomic populations in each $\ket{F,m_F}$ ground state $\rho^{\text{OP}}_{ii}$ subsequent to optical pumping and $\rho^{\text{ARP}}_{ii}$ produced after ARP. We find unique solutions for the atomic populations while also fixing the electron spin polarization $\text{p}$~\cite{suppMat}.

After spin preparation, we model the full atomic spin dynamics by solving Eq.~\eqref{eq:timeEvo} that  provides $\langle \mathcal{F}(t) \rangle$, where  $\mathcal{F}=F_{z,b}-F_{z,a}$ is the difference between the z-component of the hyperfine spin $F_{z}$ in the $b=I+1/2$ and $a=I-1/2$ manifolds. We model Faraday rotation by $ a_0+b_0\langle \mathcal{F}(t) \rangle =\theta_{F}^{\text{sim}}(t)$, where $a_0$ is a scaling  constant that we fit originating from the light-atom coupling and $b_0$ is a measured initial polarization offset of the probe beam~\cite{suppMat}.  To account for higher-order diffusion modes affecting the wall-collision rate and residual light scattering from the probe, we independently measure the non-exponential dependence of macroscopic spin decay and model as an effective time-dependent wall collision rate $\gamma_{\text{wall}}(t)$. We use fitted values $49.63$ $\mu T$ for the static magnetic field, $87.22$ kHz for the hyperfine frequency shift from buffer gas collisions, and $\{\Omega_-,\text{ }\Omega_{\pi},\text{ }\Omega_+\}=\{54.57,\text{ }35.03,\text{ }39.55 \}$ kHz extracted from Rabi oscillations driven at multiple detunings~\cite{suppMat}. 

Our model extracts $\mathcal{T}_v=107\text{ }^{\circ}$C, $P_{N_2}=190$ Torr, and p $=0.79$ from the measurements by utilizing all three Rabi lineshapes to minimize the root-mean-square error (RMSE) between the simulated $\theta_{F}^{\text{sim}}$ and the measured $\theta_{F}$~\cite{suppMat}. The inset of Fig.~\ref{fig:Figure4main}(b) displays the RMSE over variations of the vapor temperature $\mathcal{T}_{\text{v}}$ and the buffer gas pressure $\text{P}_{\text{N}_2}$, where the green circle marks the optimal values.

The dephasing rates $\gamma_{t,j} =1/T_2^{t,j}$ and nuanced features of the lineshapes predicted theoretically agree with the measured dephasing times $\gamma_{m,j} =1/T_2^{m,j}$ and lineshapes (black/green lines Fig.~\ref{fig:Figure4main}). In contrast, the corresponding weak-driving approximation predicts a total dephasing time $\gamma_{t,+}^{\text{wk}}$ for the $\sigma^+$ Rabi oscillation that underestimates $\gamma_{m,+}$ [Fig.~\ref{fig:Figure4main}(b)]. Here, $\gamma_{t,+}^{\text{wk}}$ is extracted from fitting a single decay rate to a theoretical Rabi oscillation given by an exponentially decaying sine with instantaneous decay rate  $\gamma_{\text{se}}^{\text{wk}}+\gamma_{\text{sd}}+\gamma_{\text{C}}+ \gamma_{\text{wall}}(t)$ accounting for all sources of collisional decoherence but using the weak driving approximation for the spin-exchange dephasing rate. In fact, in order for  $\gamma_{t,+}^{\text{wk}}$ to match the measured Rabi coherence would require $\mathcal{T}_{v} \rightarrow 122^{\circ}\text{C}$ to sufficiently increase the SE collision rate. This is far from the thermistor temperature $\mathcal{T}_{\text{therm}}=98^{\circ}$C, which we expect is cooler than the vapor temperature $\mathcal{T}_v$ by only a few $\degree$C due to thermal gradients across the microwave cavity. Furthermore, an electron spin polarization of p $=0.79$ is reasonable given the uncoated glass walls of our cell and the misalignment between the pump beam and the static magnetic field, and $\text{P}_{\text{N}_2}=190$ Torr is in close agreement with buffer gas pressures \{170, 180, 330\} Torr extracted independently from broadening and frequency shifts in both microwave and optical measurements~\cite{suppMat}. 

\textit{Intra-cell thermometry demonstration}---
Finally, we demonstrate the application of Rabi coherence to intra-cell thermometry [Fig.~\ref{fig:Figure4main}(c)], which is generally useful for characterizing the performance of atomic vapor sensing platforms and is particularly useful for stabilizing and optimizing the accuracy of atomic clocks~\cite{wells2014stabilizing, hao2020efforts}. In contrast, external sensors such as thermistors do not sense the actual vapor temperature. Here we cool the vapor cell by switching off the heat tape attached to the microwave cavity. While the cell cools, we record the Rabi dephasing time $T^{m,+}_2$ by generating a half-second long train of 500 $\sigma^+$ Rabi oscillations triggered every four seconds over a seven minute cooling period of 3.5 $\degree$C. We map the measured dephasing time $T^{m,+}_2\rightarrow \mathcal{T}_v$ to the intra-cell vapor temperature by varying $\mathcal{T}_v$ within the calibrated theoretical model and fitting an interpolating polynomial $\mathcal{P}$ to the simulated dephasing time $\mathcal{T}_v=\mathcal{P}(T^{t,+}_2)$. The temperature dependence of $1/T^{m,+}_2\propto n_{\text{Rb}}\sigma_{\text{se}} v^r$ arises from the atomic density, which decreases as the vapor cell cools. From a single 500 ms train of Rabi oscillations we measure a temperature sensitivity of $1.2 \text{ mK}/\sqrt{\text{Hz}}$ [Fig.~\ref{fig:Figure4main}(c) inset]. This is a competitive sensitivity with other techniques~\cite{wells2014stabilizing, salleras2009predictive}, and demonstrates Rabi coherence as a novel platform for intra-cell thermometry.

\textit{Conclusions.}--- Rabi oscillations between hyperfine manifolds in heated vapor cells exhibit nuanced time-dependence due to the SE effects on atomic dynamics, which prevents accurate evaluation and understanding of sensitivities for the increasing applications of Rabi oscillations in metrology applications. This work completes previous investigations of hyperfine coherence by using an exact numerical model and continuous QND sensing in a vapor cell to study significant deviations of SE coherence from the weak driving predictions assumed historically. We find excellent agreement with the coherence predicted by our model and Rabi oscillation measurements. Moreover, we find further agreement with the fact that we can exploit multiple hyperfine transitions to determine vapor-cell parameters and temperature dependence that are consistent with independent characterizations. This extended information that harnesses SE effects may be useful for vapor-cell characterization and intra-cell thermometry in future experiments, and the continuous Rabi measurements presented here will be an important component of working toward a use in absolute vector magnetometry~\cite{thiele2018self}.

This work was supported by DARPA through ARO grant numbers W911NF-19-1-0330 and W911NF-21-1-0127, NSF QLCI Award OMA - 2016244, ONR Grant No. N00014-17-1-2245, and NSF Grant No. PHYS 1734006.  We acknowledge helpful conversations with Yuan-Yu Jau, Christoph Affolderbach, Vladislav Gerginov and Michaela Ellmeier, and technical expertise from Thanmay Sunil Menon, Yolanda Duerst, and Felix Vietmeyer.


%

\end{document}


\onecolumngrid

\title{Supplementary Material for \lq\lq Coherence of Rabi oscillations with spin exchange"}

\author{C.~Kiehl}\email{christopher.kiehl@colorado.edu}
\affiliation{JILA, National Institute of Standards and Technology and University of Colorado, and
Department of Physics, University of Colorado, Boulder, Colorado 80309, USA}
\author{D.~Wagner} 
\affiliation{JILA, National Institute of Standards and Technology and University of Colorado, and
Department of Physics, University of Colorado, Boulder, Colorado 80309, USA}
\author{T.-W.~Hsu} 
\affiliation{JILA, National Institute of Standards and Technology and University of Colorado, and
Department of Physics, University of Colorado, Boulder, Colorado 80309, USA}
\author{S.~Knappe} 
\affiliation{Mechanical Engineering, University of Colorado, Boulder, Colorado 80309, USA}
\affiliation{FieldLine Inc., Boulder CO 80301, USA}
\author{C.~A.~Regal} 
\affiliation{JILA, National Institute of Standards and Technology and University of Colorado, and
Department of Physics, University of Colorado, Boulder, Colorado 80309, USA}
\author{T.~Thiele} 
\altaffiliation[Current address: ]{Zurich Instuments AG, Technoparkstrasse 1, 8005 Zurich, Switzerland}
\affiliation{JILA, National Institute of Standards and Technology and University of Colorado, and
Department of Physics, University of Colorado, Boulder, Colorado 80309, USA}

\maketitle
\tableofcontents

\section{Symbol Definitions}
A list of symbol definitions in the main text is listed in Table~\ref{symbDefs}.
\begin{table*}
\caption{\label{symbDefs} Symbol definitions}
\begin{ruledtabular}
\begin{tabular}{c c}
Symbol & Definition\\[1mm]

\colrule
\\ [-.3em]
$\Gamma$&\text{$=n_{a}\sigma_i v^r$ is the atomic collision rate characterized by a collision cross section (e.g. $\sigma_{\text{se}}$ for SE Rb-Rb)}\\[2mm]
$\Gamma_{D}$& $=D \pi^2/(l_x^2+l_y^2+l_z^2)$ is the fundamental decay mode for diffusion into cell walls\\[2mm]
$\Gamma_{C}$& Carver rate\\[2mm]
$\gamma$& $1/T_2$\\[2mm]
$\gamma_{\text{se}}$&contribution to $\gamma$ due only to SE collisions\\[2mm]
$\gamma_{\text{sd}}$&contribution to $\gamma$ due only to SD collisions ($\gamma_{\text{sd}}\approx \Gamma_{\text{sd}}^{\text{(RbRb)}}+\Gamma_{\text{sd}}^{\text{(RbN}_2\text{)}}$)\\[2mm]
$\gamma_{\text{C}}$& $=\Gamma_{\text{C}}\eta_I^2 [I]/8$ is the contribution to $\gamma$ due only to Carver dephasing\\[2mm]
$\gamma_{\text{wall}}$& $=\Gamma_{\text{D}}$ is the contribution to $\gamma$ due only to wall collisions assuming the fundamental diffusion mode\\[2mm]
$\gamma_{\text{se}}^{\text{wk}}$& weak driving aproximation for $\gamma_{\text{se}}$\\[2mm]
$\gamma_{t,j}$& Theoretical $\gamma$ for the $j\in\{\pm,\pi \}$ Rabi oscillation \\[2mm]
$\gamma_{m,j}$& Measured $\gamma$ for the $j\in\{\pm,\pi \}$ Rabi oscillation\\[2mm]
$\gamma_{t,+}^{\text{wk}}$&$\gamma_{t,+}$ due to all collision processes predicted if $\gamma_{\text{se}}$ is replaced by $\gamma_{\text{se}}^{\text{wk}}$\\[2mm]

\end{tabular}
\end{ruledtabular}
\end{table*}
\section{Atom-microwave coupled Hamiltonian}
\subsection{Derivation and making RWA}
The full Hamiltonian to drive Rabi oscillations between the hyperfine ground states of alkali atoms is given by \begin{align}
\begin{split}
\label{eq:SMfullHam}
    H(t)=(A_{\text{hfs}}+h\delta \nu_{\text{hfs}}/2)\mathbf{S}\cdot \mathbf{I}+\mu_{B}(g_s \mathbf{S}+g_I \mathbf{I})\cdot(B^{\text{dc}}\hat{z}+\mathbf{\mathcal{B}}^{\mu}(t))
    =H_{\text{hfs}}+\delta H_{\text{hfs}}+H_{\text{dc}}+H_{\mu}(t)
\end{split}
\end{align}
where $A_{\text{hfs}}$ ($=h\cdot 3.417$ GHz for $^{87}$Rb) is the ground state magnetic dipole constant, $\mathbf{S}$ and $\mathbf{I}$ are the electron and nuclear spin operators respectively with corresponding g-factors $g_s$ and $g_I$, $B^{\text{dc}}$ is the strength of a DC magnetic field, which we may define without loss of generality to point along $\hat{z}$, and the microwave field can be expressed as~\cite{thiele2018self}
\begin{equation}
    \mathbf{\mathcal{B}}^{\mu}(t)=\sum_{i=x,y,z}\mathcal{B}_i\cos(\omega_{\mu}t+\phi_i)\hat{e}_i.
\end{equation}
where $\nu_{\mu}=\omega_{\mu}/2 \pi$ is the microwave frequency. We work in the $\ket{F,m_F}$ basis, which diagonalizes the hyperfine structure $A_{\text{hfs}}\mathbf{S}\cdot\mathbf{I}$ (but not $H_{\text{dc}}$ nor $H_{\mu}$), where $\mathbf{F}=\mathbf{S}+\mathbf{I}$ is the total angular momentum operator. Here $F=I\pm1/2$ defines the two hyperfine levels. The  hyperfine frequency shift $\delta \nu_{\text{hfs}}$ is due to perturbations of the electron cloud during $\text{N}_2$ buffer gas collisions that on average increases the hyperfine coupling constant.

Due to the large hyperfine constant $A_{\text{hfs}}>>\mu_B B^{\text{dc}}$ ($B^{\text{dc}}\approx 50$ $\mu$T), the exact solution to $\dot{\rho}=-\frac{i}{\hbar}[H(t),\rho]$ results in rapidly oscillating hyperfine coherences that increases computational time. We minimize this computational time by making the rotating wave approximation (RWA) into a time-independent frame, which removes the large energy scale caused by $A_{\text{hfs}}$. To make the RWA, we define the diagonalized unitary operator $\mathcal{U}$ with diagonal elements  $\mathcal{U}_{ii}=1$ ($\mathcal{U}_{ii}=e^{\text{-}i\omega_{\mu} t}$) for states in the $F=1$ ($F=2$) manifold, and make the transformation
\begin{equation}\tilde{H}(t) =\mathcal{U}^{\dagger}H(t)\mathcal{U}-i\mathcal{U}^{\dagger}\frac{d\mathcal{U}}{dt} \end{equation}

Next, we eliminate the counter rotating terms by numerically averaging $\tilde{H}(t)$ with time-steps $\Delta t= 1/(\nu_\mu N_{\text{ave}})$, where $N_{\text{ave}}$ is the number of averages
\begin{equation}
 \tilde{H}=\frac{1}{N_{\text{ave}}}\sum_{k=0}^{N_{\text{ave}}-1} \tilde{H}(k \Delta t)
\end{equation}
such that high-frequency terms $\propto e^{\pm i n \omega_{\mu}t}$ oscillating at multiple integers of $\nu_{\mu}$ are eliminated. Here we use $N_{\text{ave}}=4$, where $e^{\pm i n \omega_{\mu}t}$ terms in this rotated frame are efficiently averaged away using time-steps $\Delta t$. By doing this we have also eliminated coupling between the hyperfine levels arising from off-diagonal elements in $H_{\text{dc}}$: $\bra{2I-1,m}H_{\text{dc}}\ket{2I+1,m^{\prime}}$, which would overwise quickly oscillate and average out. We have also eliminated microwave coupling terms between magnetic sublevels within each hyperfine manifold $\bra{F,m}H_{\mu}\ket{F,m^{\prime}}$ that are otherwise far-detuned ($\Delta_{\mu w}\approx 2A_{\text{hfs}}/h$) from the single-photon transition resonance.

In terms of the microwave field, the Rabi rate for a given microwave transition is $\Omega_{j}=\mu_{j} \mathcal{B}_{j}/h$ where $j=\pm,\pi$ and $\mathcal{B}_{\pi}=\mathcal{B}_z$ and $\mathcal{B}_{\pm}=\frac{\pm\mathcal{B}_x + i\mathcal{B}_y}{\sqrt{2}}$ are the spherical microwave components~\cite{thiele2018self}. The transition dipole matrix elements $\mu_{F,m_F,F^{\prime},m_F^{\prime}}$ are defined by~\cite{thiele2018self}
\begin{align}
\begin{split}
\mu_{F,m_F,F^{\prime},m_F}=\mu_{\pi}
=\frac{\mu_B}{2\hbar}\bra{F^{\prime},m_F}(g_sS_z+g_II_z)\ket{F,m_F}
\end{split}
\end{align}
\begin{align} 
\begin{split}
\mu_{F,m_F,F^{\prime},m_F\pm1}=\mu_{\pm}
=\frac{\mu_B}{2\sqrt{2}\hbar}\bra{F^{\prime},m_F\pm1}(g_sS_{\pm}+g_II_{\pm})\ket{F,m_F}
\end{split}
\end{align}
where $S_{\pm}=S_x\pm iS_y$ and $I_{\pm}=I_x\pm iI_y$.

\subsection{Making the RWA on the decoherence operators}

We must also place the decoherence operators in Eq.~\eqref{eq:SMfullHam}, e.g. $\Gamma_{\text{se}}\big(\phi(1+4\langle \mathbf{S} \rangle \cdot \mathbf{S} )-\rho \big)$ where $\phi=\rho/4-\mathbf{S}\cdot \rho \mathbf{S}$, in the rotating frame and make the RWA accordingly. To do this we first put the spin matrices in the rotating frame $\tilde{\mathbf{S}}(t)=\mathcal{U}^{\dagger}\mathbf{S} \mathcal{U}$ and then numerically average the decoherence operators with time-steps $\Delta t= 1/(\nu_\mu N_{\text{ave}})$, where $N_{\text{ave}}$ is the number of averages. We cannot average the spin-matrices individually since the decoherence operators contain higher-order products like $\tilde{S}_x(t)\rho(t) \tilde{S}_x(t)$ that contain nontrivial cancellation of counter-rotating terms before averaging. Instead, we calculate
\begin{align}
\begin{split}
  \tilde{S}_x(t)\rho(t) \tilde{S}_x(t)  \rightarrow  &\frac{1}{N_{\text{ave}}}\sum_{k=0}^{N_{\text{ave}}-1} \tilde{S}_x(k \Delta t) \rho (t) \tilde{S}_x(k \Delta t)
\end{split}
\end{align}
such that high-frequency terms $\propto e^{\pm i n \omega_{\mu}t}$ oscillating at multiple integers of $\nu_{\mu}$ are eliminated. Note that since we are using the time-independent Hamiltonian $\tilde{H}$, $\rho(t)$ in the rotated frame contains no counter-rotating terms and is left fixed at time $t$ during the averaging. Here we again use $N_{\text{ave}}=4$, where $e^{\pm i n \omega_{\mu}t}$ terms in this rotated frame are efficiently averaged away using time-steps $\Delta t$. Using more averaging terms only increases computation time. 

\subsection{Spin matrices}
Here we write the electron and nuclear spin operators $\mathbf{S}=\{S_x,S_y,S_z \}$ and $\mathbf{I}=\{I_x, I_y, I_z \}$ in the $\ket{F,m_F}$ basis. This is done by first writing $\mathbf{S}$ and $\mathbf{I}$ in the $\ket{S,I,m_s,m_I}$ eigenbasis, defined by $S_z\ket{S,m_s}=m_s\ket{S,m_s}$ and $S_{\pm}\ket{S,m_s}=(S_x\pm iS_y)\ket{S,m_s}=\sqrt{S(S+1)-m_s(m_s\pm1)}\ket{S,m_s\pm1}$ (with the same corresponding definitions for $I_z$ and $I_{\pm}$), and making a basis transformation by using Clebsch-Gordan coefficients $c_{\{F,m_F,m_s,m_I \}}$ defined by $\ket{F,m_F}=\sum c_{\{F,m_F,m_s,m_I \}} \ket{m_s,m_I}$. Using these coefficients over the different combinations of the $n=4I+2$ basis states we write a transformation matrix $\mathcal{G}$
\[
  \mathcal{G} =
  \begin{bmatrix}
    c_{\{(F,m_F)^{(1)},(m_s,m_I)^{(1)} \}}&\hdots & c_{\{(F,m_F)^{(n)},(m_s,m_I)^{(1)}\}}\\
    \vdots& \ddots & \\
    & &\\
    c_{\{(F,m_F)^{(1)},(m_s,m_I)^{(n)}\}} & &c_{\{(F,m_F)^{(n)},(m_s,m_I)^{(n)} \}}
  \end{bmatrix}
\]
such that $\mathcal{M}\rightarrow \mathcal{G}^\dagger \mathcal{M} \mathcal{G}$ achieves the $\ket{m_s,m_I}\rightarrow \ket{F,m_F}$ transformation for an arbitrary operator $\mathcal{M}$. We specifically show $\mathcal{G}$ in Table \ref{clesbschG} for the $F=1$ and $F=2$ ground states of $^{87}$Rb, 
from which we use to calculate the electron and nuclear spin matrices in the $\ket{F,m_F}$ basis (Tables \ref{elecSpin} and \ref{nucSpin}).
\begin{table}
\caption{\label{clesbschG}%
Transformation matrix $\mathcal{G}$ for making the basis transformation $\ket{m_s,m_I}\rightarrow \ket{F,m_F}$ of the $F=1$ and $F=2$ ground states for $^{87}$Rb.}
\begin{ruledtabular}
\begin{tabular}{c c c c c c c c c}
\textrm{\quad} &
\textrm{\scriptsize$ \ket{1,1}$} &\textrm{\scriptsize$ \ket{1,0}$} &\textrm{\scriptsize$ \ket{1,-1}$} &\textrm{\scriptsize$ \ket{2,2}$} &\textrm{\scriptsize$ \ket{2,1}$} &\textrm{\scriptsize$ \ket{2,0}$} &\textrm{\scriptsize$ \ket{2,-1}$} &\textrm{\scriptsize$ \ket{2,-2}$}\\[1mm]
\colrule
\\ [-.3em]
\textrm{\scriptsize$ \bra{-\frac{1}{2},\frac{1}{2}}$} &0 &$-\frac{1}{\sqrt{2}}$ &0 &0 &0 &$\frac{1}{\sqrt{2}}$&0&0\\[2mm]
\textrm{\scriptsize$ \bra{-\frac{1}{2},-\frac{1}{2}}$} &0 &0 &$-\frac{1}{\sqrt{2}}$ &0 &0 &0&$\sqrt{\frac{3}{2}}$&0\\[2mm]
\textrm{\scriptsize$ \bra{-\frac{1}{2},-\frac{3}{2}}$} &0 &0 &0 &0 &0 &0&0&1\\[2mm]
\textrm{\scriptsize$ \bra{\frac{1}{2},\frac{3}{2}}$} &0 &0 &0 &1 &0 &0&0&0\\[2mm]
\textrm{\scriptsize$ \bra{\frac{1}{2},\frac{1}{2}}$} &$\frac{1}{2}$ &0 &0 &0 &$\frac{\sqrt{3}}{2}$ &0&0&0\\[2mm]
\textrm{\scriptsize$ \bra{\frac{1}{2},-\frac{1}{2}}$} &0 &$\frac{1}{\sqrt{2}}$ &0 &0 &0 &$\frac{1}{\sqrt{2}}$&0&0\\[2mm]
\textrm{\scriptsize$ \bra{\frac{1}{2},-\frac{3}{2}}$} &0 &0 &$\frac{\sqrt{3}}{2}$ &0 &0 &0&$\frac{1}{2}$&0\\[2mm]
\textrm{\scriptsize$ \bra{-\frac{1}{2},\frac{3}{2}}$} &$-\frac{\sqrt{3}}{2}$ &0 &0 &0 &$\frac{1}{2}$ &0&0&0\\[2mm]
\end{tabular}
\end{ruledtabular}
\end{table}

\section{Details on atomic collisions}
\subsection{Collision rates}
As noted in the main text, a dephasing rate $\gamma$ due to a collisional process is proportional to the corresponding collision rate $\Gamma$. The collision rate is characterized by a cross section $\sigma$ with the form $\gamma\propto \Gamma \equiv n_{a}\sigma v^{r}$, where $v^r$ is the average relative velocity, and the atomic density $n_{a}$ $(n_{\text{buff}})$ for alkali collisions (buffer gas collisions)~\cite{allred2002high-sensitivity}. Specifically for $n_a=n_{\text{Rb}}$ and $n_{\text{buff}}=n_{\text{N}_2}$, we have from the ideal gas law $n_{\text{Rb}}=\text{P}_{\text{Rb}}/k_B \mathcal{T}_v$ and $n_{\text{N}_2}=\text{P}_{\text{N}_2}/k_B \mathcal{T}_v$, where $\text{P}_{\text{Rb}}$ is given by the vapor-pressure model for the liquid Rb phase~\cite{nesmeianov1963vapor}. Typically these cross sections are reported as the parameter $\kappa=\sigma v^r$, where $v^r=\sqrt{8k_B \mathcal{T}_v/\pi m_{\mu}}$ is the mean atomic speed between the colliding pair and $m_{\mu}$ is the reduced mass of the colliding pair. Sometimes a slowing down factor $q$ is incorporated into the reported value of $\kappa$. Below we list some of the collisional cross-sections reported in the literature with  ($\star$) marking the values used in this work, and explicitly note if $q$ is incorporated into the reported $\kappa$:

\underline{$\Gamma_{\text{sd(RbRb)}}$}:
\begin{align*}
   \kappa_{\text{sd(RbRb)}}=8.11 \pm 0.33\times 10^{-19} \text{ m}^3\text{s}^{-1}\text{ \cite{chupp1989laser}}\\
    \Rightarrow  \sigma_{\text{sd(RbRb)}}=1.77\pm0.07\times 10^{-21}\text{ m}^2 \text{ }(\star)
\end{align*}

assuming $\mathcal{T}_v=150^{\circ}$C during the measurement~\cite{chupp1989laser}.
\begin{align*}
  \sigma_{\text{sd(RbRb)}}=1.6\pm0.2\times 10^{-21}\text{ m}^2 \text{ ~\cite{knize1989spin}}
\end{align*}

\begin{align*}
   \kappa_{\text{sd(RbRb)}}=3.9 \pm 0.4\times 10^{-20} \text{ m}^3\text{s}^{-1}\text{ \cite{baranga1998polarization}}\\
    \Rightarrow \sigma_{\text{sd(RbRb)}}=9.2\pm0.9\times 10^{-22}\text{ m}^2 
\end{align*}

assuming $\mathcal{T}_v=150^{\circ}$C used in~\cite{baranga1998polarization} and that $\kappa_{\text{sd(RbRb)}}=\sigma_{\text{sd(RbRb)}}v^r/q$, where $q=10.8$ is a slowing down-factor used for calculating decay rates valid for low spin-polarization.

\underline{$\Gamma_{\text{sd(Rb}N_2\text{)}}$}:
\begin{align*}
   \kappa_{\text{sd(Rb}N_2\text{)}}=9.38 \pm 0.22\times 10^{-24} \text{ m}^3\text{s}^{-1}\text{ \cite{chupp1989laser}}\\
    \Rightarrow \sigma_{\text{sd(Rb}N_2\text{)}}=1.44\pm0.03\times 10^{-26}\text{ m}^2 \text{ }(\star)
\end{align*}

assuming $\mathcal{T}_v=150^{\circ}$C~\cite{chupp1989laser}.

\underline{$\Gamma_{\text{se(RbRb)}}$}:
\begin{align*}
  \sigma_{\text{se(RbRb)}}=1.9\pm0.2\times 10^{-18}\text{ m}^2 \text{ \cite{gibbs1967spin}} \text{ }(\star)
\end{align*}
where this same value is cited in~\cite{sheng2013subfemtotesla}.

\subsection{SE weak-driving approximation}
To analytically describe hyperfine decoherence due to  SE collisions for weak driving, the typical path \cite{vanier1989quantum} is to first write $
d\rho/dt=\hat{\Gamma}_{ex}(\rho)=\big[\phi(1+4\langle \mathbf{S} \rangle \cdot \mathbf{S})-\rho\big]T_{ex}^{-1}$ in the approximation such that $\rho$ is diagonal except for the coherence $\rho_{ij}$ that corresponds to the driven transition, and assuming fixed populations given by a spin-temperature distribution $\rho_{ii}=\rho^{(p)}_{ii}\propto e^{-\beta(p)m_i}$, which we show for the $\sigma^+$ transition below: 
\begin{equation}\rho^{\text{wk}}_{+}\rightarrow\left[\begin{array}{*8{C{1em}}}
   \rho_{11}&0&0&\rho_{14}&0&0&0&0\\[1mm]
   0&\rho_{22}&0&0&0&0&0&0\\[1mm]
   0&0&\rho_{33}&0&0&0&0&0\\[1mm]
  \rho_{41}&0&0&\rho_{44}&0&0&0&0\\[1mm]
   0&0&0&0&\rho_{55}&0&0&0\\[1mm]
   0&0&0&0&0&\rho_{66}&0&0\\[1mm]
   0&0&0&0&0&0&\rho_{77}&0\\[1mm]
   0&0&0&0&0&0&0&\rho_{88}
   \end{array}\right]
\end{equation}
where we use the index notation $\ket{F,m_F}\rightarrow i=4F-m_F-2$, where for example $\bra{1,1}\rho\ket{2,2}\rightarrow \rho_{14}$.
By evaluating $\hat{\Gamma}_{ex}(\rho_{+}^{\text{wk}})$, we obtain an exponential decay equation for the $\sigma^+$ coherence 
\begin{align}
\begin{split} \label{eq:endR}
   \frac{\text{d}\rho_{41}}{\text{dt}}&=-\gamma^{\text{wk}}_{\text{se,+}}\rho_{41}=\\ \frac{\rho_{41}}{T_{ex}}&\bigg[\frac{7}{8}(\rho_{44}-1)+\frac{\rho_{11}}{2}+\frac{\rho_{33}-\rho_{77}}{16}+\frac{\rho_{55}}{4}-\frac{\rho_{88}}{8}\bigg]\\
   &\rightarrow-\frac{\rho_{41}}{T_{ex}}\frac{(1-p)(11+5p^2)}{16(1+p^2)}\text{\quad \quad(ST Distr.)}
   \end{split}
\end{align}
By assuming fixed atomic populations, we obtain an exponential decay rate of the $\rho_{41}$ hyperfine coherence equal to the rate reported by Y. Y. Jau et al. for weak microwave driving~\cite{jau2004intense}. This same analysis for the $\sigma^-$ transition gives
\begin{align}
\begin{split} 
   \frac{\text{d}\rho_{61}}{\text{dt}}&=\\ \frac{\rho_{61}}{T_{ex}}&\bigg[3\frac{\rho_{11}}{32}+\frac{\rho_{22}-\rho_{33}}{16}+\frac{\rho_{88}-\rho_{44}}{8}+\frac{\rho_{55}}{32}-\frac{3}{4}+\frac{\rho_{66}+\rho_{77}}{16}\bigg]\\
   &\rightarrow-\frac{\rho_{61}}{T_{ex}}\frac{23+p(3+5p(5+p))}{32(1+p^2)} \text{\quad (ST Distr.)}\\
   &\rightarrow-\frac{\rho_{61}}{T_{ex}}\frac{92+p(5+p(79-p-7p^2))}{128(1+p^2)} \text{   (ARP)}
   \end{split}
\end{align}

Here the label \lq ARP\rq\text{ }refers to the initial atomic state after adiabatic rapid passage, where the atomic populations in the $\sigma^+$ transition are switched ($\rho_{11}\leftrightarrow {\rho_{44}}$) within a ST distribution. In principle the RWA must be correctly applied to $\hat{\Gamma}_{ex}(\rho_{j}^{\text{wk}})$ as outlined in a previous section. This, however, is only necessary for $j=\pi$, whereas the weak-driving approximation for the $j=\pm$ dephasing rates remain unchanged after the RWA. For the $\pi$ transition this analysis with RWA gives
\begin{align}
\begin{split}
   \frac{\text{d}\rho_{15}}{\text{dt}}&=\\ -\frac{\rho_{15}}{T_{ex}}&\frac{1}{32}\bigg[26-6\rho_{11}-3(\rho_{22}+2(\rho_{44}+\rho_{55})+\rho_{66})\bigg]\\
   &\rightarrow-\frac{\rho_{51}}{T_{ex}}\frac{23-6p+23p^2}{32(1+p^2))} \text{\quad \quad(ST Distr.)}\\
   &\rightarrow-\frac{\rho_{51}}{T_{ex}}\frac{23-6p+23p^2}{32(1+p^2))} \text{\quad \quad \quad (ARP)}
   \end{split}
\end{align}

For context, other works involving atomic clocks \cite{vanier1989quantum} assume that optical pumping simply depopulates the lower hyperfine manifold ($F=1$) and leaves the upper manifold ($F=2$) equally populated. While this is physically not the situation in this work, the equations simplify to pure exponential decay equations that are unaffected by the RWA.
\begin{align}
\begin{split}
  \frac{d\rho_{62}}{dt}&=-\frac{5}{8}\rho_{62} \text{\quad \quad \quad($\ket{1,0}\rightarrow \ket{2,0}$)}\\
  \frac{d\rho_{51}}{dt}&=-\frac{23}{32}\rho_{51
  } \text{  \quad \quad($\ket{1,1}\rightarrow \ket{2,1}$)}
   \end{split}
\end{align}

\section{Modeling Experiment}
\subsection{Faraday Rotation}
Faraday rotation of linearly-polarized light passing through atomic vapor is given by~\cite{seltzer2008developments}
\begin{equation}
    \theta_F=\frac{\pi \nu l}{c}(n_+(\nu)-n_-(\nu))
\end{equation}
where $n_{\pm}(\nu)$ is the index of refraction for circularly-polarized light at optical frequency $\nu$. Spin-polarized atomic vapors exhibit birefringence $n_{+}(\nu)\neq n_-(\nu)$. This birefringence can be expressed through the bulk atomic polarizability~\cite{dressler1996theory} \begin{equation}
    n_{\pm}(\nu)=\sqrt{1+\alpha^{\pm}(\nu)}\approx 1+\frac{1}{2}\alpha^{\pm}(\nu)
\end{equation}
where the bulk atomic polarizability is expressed as
\begin{equation}
\alpha^q(\nu)=\sum_{F,m_F} \rho(F,m_F)\alpha^q_{F,m_F}(\nu)
\end{equation}
that is weighted with the atomic populations $\rho(F,m_F)$ and $ \alpha^{q}_{F,m_F}(\nu)$ is given by
\begin{align}
    \begin{split}
    \alpha^{q}_{F,m_F}(\nu)=\frac{n_{a}}{\epsilon_0 h}\sum_{F',m_{F'}}&|\bra{F,m_F}e r_q\ket{F',m_{F'}}|^2\times \mathcal{D}(\nu-\nu_{\text{tr}})
    \end{split}
\end{align}
Here $\bra{F,m_F}e r_q\ket{F',m_{F'}}$ is the transition dipole matrix element with $q=\pm,0$ denoting the spherical component, $n_a$ is the alkali atomic density, $\epsilon_0$ is the permitivity of free space, $\mathcal{D}(\nu-\nu_{\text{tr}})=(\nu-\nu_{\text{tr}})/[(\nu-\nu_{\text{tr}})^2+(\gamma_{\text{D2}}/2)^2]$ is a dispersive lineshape with FWHM $\gamma_{\text{D2}}$, and $\nu-\nu_{\text{tr}}$ is the optical detuning for the transition $\text{tr}=\ket{F,m_F}\rightarrow \ket{F',m_{F'}}$.
The transition dipole matrix element is written as~\cite{steckrubidium}
\begin{align}
\begin{split}
    |\bra{F,m_F}e &r_q\ket{F',m_{F'}}|^2=|\bra{F|}e\mathbf{r}\ket{|F'}|^2 
    \times (2F+1) \begin{pmatrix}
  F' & 1 & F \\
  m'_{F} & q & -m_F 
\end{pmatrix}^2
    \end{split}
\end{align}
where
\begin{align}
\begin{split}
   |\bra{F|}e\textbf{r}\ket{|F'}|^2&=|\bra{J|}e\mathbf{r}\ket{|J'}|^2 (2F'+1)(2J+1)\times  \begin{Bmatrix}
  J & J' & 1 \\
  F' & F & I 
\end{Bmatrix}^2
    \end{split}
\end{align}
and $|\bra{J|}e\mathbf{r}\ket{|J'}|^2=3h\epsilon_0r_ec^2f_{osc}/2\pi\nu_0$ with $r_e=2.82\times 10^{-15}$ m is the classical electron radius, $f_{osc}$ is the oscillator strength for the optical transition, and $\nu_0$ is the optical transition frequency. 

Putting this all together for the case of $^{87}$Rb gives
\begin{align}
\begin{split}
   \theta_F&=\frac{3c r_e f_{osc}n_{Rb}l}{2I+1}\sum_{F,m_F}\rho(F,m_F)(2F+1)\\
    &\times \sum_{F',m'_F}(2J+1)(2F'+1)\begin{Bmatrix}
  J & J' & 1 \\
  F' & F & I 
\end{Bmatrix}^2 \\
&\times\bigg[\mathcal{D}(\nu-\nu_a)\begin{pmatrix}
  F' & 1 & F \\
  m_F+1 & -1 & -m_F 
\end{pmatrix}^2-\mathcal{D}(\nu-\nu_b)\begin{pmatrix}
  F' & 1 & F \\
  m_F-1 & 1 & -m_F 
\end{pmatrix}^2 \bigg]
    \end{split}
\end{align}
where we approximate $\nu/\nu_0\approx 1$.
Here $a\rightarrow I-1/2$ and $b \rightarrow I+1/2$ denote the different optical detunings from the $F=1,2$ manifolds. For large detunings $|\nu-(\nu_a-\nu_b)/2| \gg |\nu_a-\nu_b|=6.8$ GHz, which is the case in this work, we may approximate $\mathcal{D}(\nu-\nu_{\text{a}})\approx \mathcal{D}(\nu-\nu_{\text{b}})$~\cite{kong2020measurement}. Then, for atomic ensembles prepared in specfic $\ket{F,m_F}$ states
\begin{equation}
    \theta_F=\frac{c r_e f_{D1}n_{Rb}l}{2I+1}\mathcal{D}(\nu-\nu_{D1})\begin{cases}
    +m_F \text{ (F = 1)}\\
    -m_F \text{ (F = 2)}
    \end{cases}
\end{equation}
\begin{equation}
    \theta_F=\frac{c r_e f_{D2}n_{Rb}l}{2(2I+1)}\mathcal{D}(\nu-\nu_{D2})\begin{cases}
    -m_F \text{ (F = 1)}\\
    +m_F \text{ (F = 2)}
    \end{cases}
\end{equation}
for $D_1$ and $D_2$ transitions respectively ($D_1$ case is consistent with Refs.~\cite{shah2010high,kong2020measurement}).
This implies that we may write for the $D_2$ case
\begin{equation}
\label{eq:faraday}
    \theta_F\propto \langle \mathcal{F} \rangle = \langle F_{z,b} - F_{z,a} \rangle
\end{equation}
where $\langle F_{z,a}\rangle$ and $\langle  F_{z,b}\rangle$ denote the expectation value of the hyperfine spin for the $F =1$ and $F=2$ manifolds respectively.
\begin{figure*}[tbh]
\includegraphics[width=0.99\textwidth]{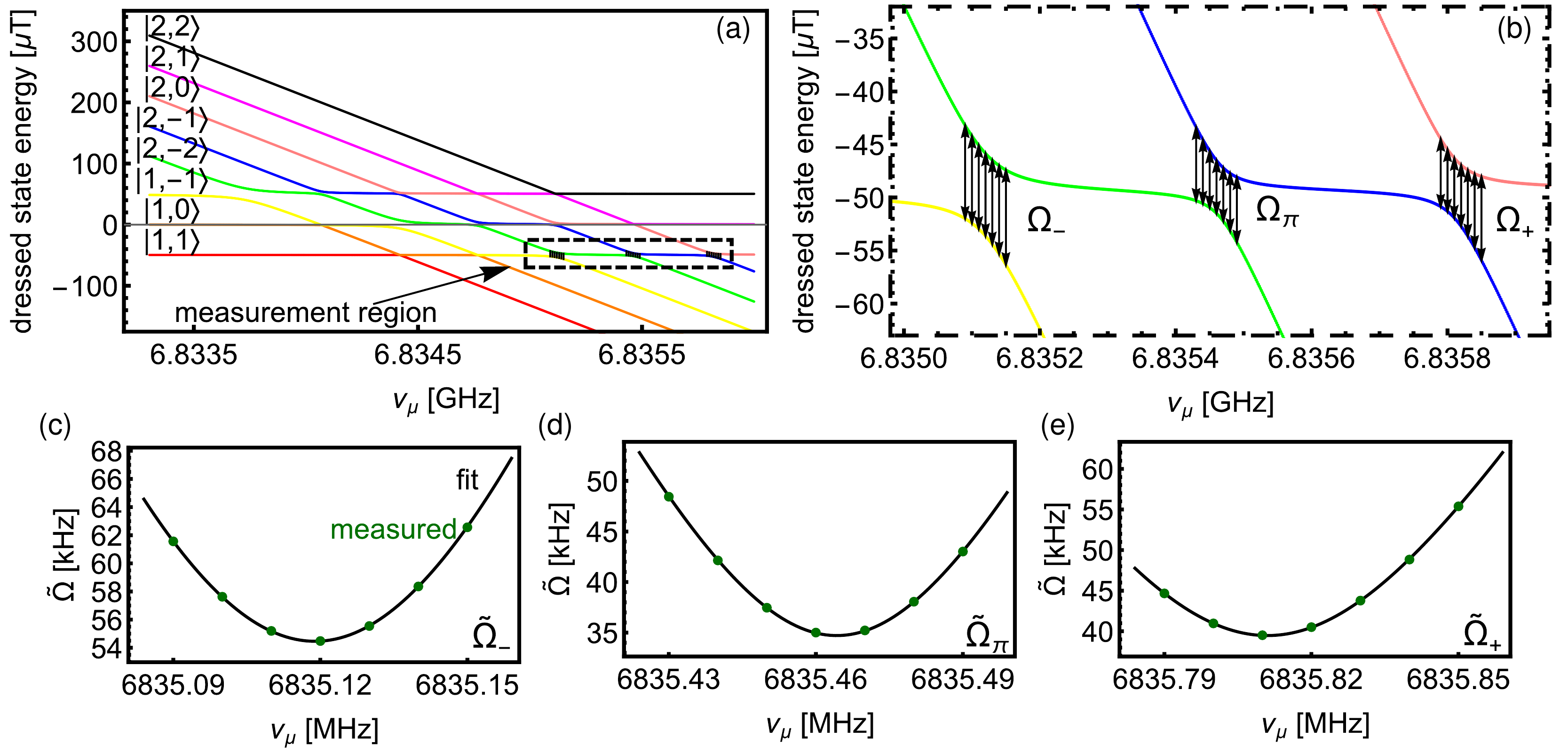}
\caption{Extraction of $\delta \nu_{hfs}$, $|\vec{B}|$, and Rabi frequencies. (a) Eigenvalues (dressed states) of the total microwave-coupled Hamiltonian. Units are reported in $\mu T=[E/\mu_B]$. (b) Enlargement of the measurement region of (a) where the specific microwave frequencies for each microwave transition are shown. (c-e) Fits (black) using the dressed states of the full Hamiltonian overlayed with measured generalized Rabi frequencies (green).}
\label{fig:smRabiACStark}
\end{figure*}
\subsection{Extraction of Rabi frequencies}
To accurately model Rabi oscillation measurements requires knowing the Rabi frequencies, the magnetic field to calculate Zeeman shifts, and the hyperfine frequency shift $\delta \nu_{hfs}$ arising from buffer gas collisions. We extract these parameters by fitting to generalized Rabi frequency measurements $\tilde{\Omega}_j$ with $j\in\{\pm,\pi\}$ at seven different detunings near resonance for each transition. Quantitatively, these generalized Rabi oscillations are described by the eigenvalues $\lambda_r$ with $r\in \{1,...,8\}$ of the full time-independent Hamiltonian $\tilde{H}$. We show in Fig.~\ref{fig:smRabiACStark}(a) these eigenvalues, or rather dressed-states, as a function of microwave frequency $\nu_{\mu}$ for the fitted parameters [Fig.~\ref{fig:smRabiACStark}(c-e)] $\Omega_-= 54.565\text{ kHz}$, $\Omega_{\pi}=35.025\text{ kHz}$, $\Omega_+=39.552\text{ kHz}$, $|\vec{B}|=49.626\text{ }\mu T$, $\delta \nu_{hfs}=87.219 \text{ kHz}$. Here, the generalized Rabi frequency is given by difference between two dressed at an anti-crossing as shown in Fig.~\ref{fig:smRabiACStark}(b). 

\subsection{Spin decay measurements and modeling wall-collision decoherence}
\begin{figure}[tbh]
\includegraphics[width=0.49\textwidth]{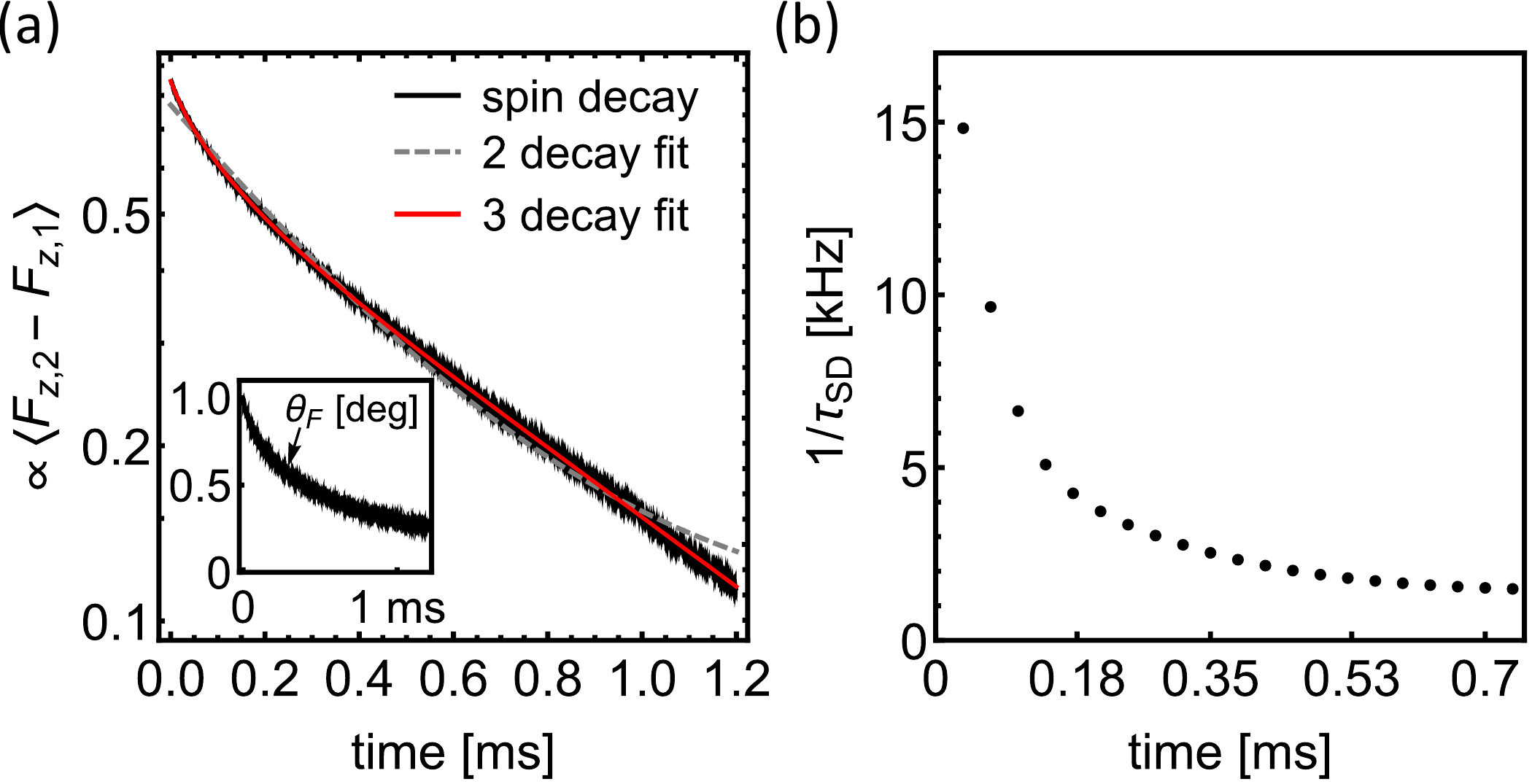}
\caption{\label{fig:Systematics} Spin decay from wall collisions (a) Spin decay after optical pumping. Inset shows Faraday rotation $\theta_F\propto a_0+b0 \langle \mathcal{F} \rangle$ when $\vec{B}^{dc}$ is parallel with the pump beam. We use a decay fitting function $a0+\sum_{i}A_ie^{-t/\tau_i}$ and fit to the Faraday rotation data of the inset. The main log-plot displays the spin decay ($\propto \mathcal{F}$) with the constant offset $a_0$ removed to show the deviation from a pure exponential decay due to higher-order wall-collision modes. (b) Instantaneous spin-decay rate obtained by fitting a single exponential decay to 35 $\mu$s sections of the \lq 3 decay' fit. }
\label{fig:smWallColl}
\end{figure}

To properly model Rabi oscillation measurements requires correctly modeling the time-dependent spin decay. To measure spin decay, we align $\vec{B}^{\text{dc}}$ with our pump and probe beam such that no Larmor precession occurs. Then we measure the decay of the probe Faraday rotation $\theta_F(t)$ after turning off the pump beam as shown in Fig.~\ref{fig:smWallColl}(a). In this configuration, the Faraday angle senses the magnitude of the macroscopic spin state. We fit $\theta_F(t)=a0+\sum_{i}A_ie^{-t/\tau_i}$ to the spin decay [Fig.~\ref{fig:smWallColl}(a)], where we obtain a good fit using at least 3 decay rates. We attribute this time-dependent spin-decay to be primarily due to higher-order wall-collision modes and residual light scattering from the probe beam. Because the pump beam is slightly focused into the vapor cell to achieve a higher pumping rate, we expect the presence of higher-order diffusion modes in our measurements. Since we cannot easily distinguish decoherence due to wall collisions from light scattering, we model spin decay as an effective time-dependent wall collision decay rate [Fig.~\ref{fig:smWallColl}(b)].

Spin decay from wall collisions occurs due to large time-dependent electric and magnetic fields experienced by the atom when colliding with the cell wall surface that effectively randomizes the atomic spin. The diffusion of the atomic spin polarization $p(x,y,z,t)$ spreads throughout the vapor cell according to the diffusion equation $\partial p/\partial t=D\nabla^2 p$, where $D=D_0 P_0/P_{\text{buff}}$ is a diffusion constant that is inversely proportional to the buffer-gas pressure, and $P_0=1$ atm. The boundary condition $p=0$ at the cell walls is assumed, which leads to solutions due to the diffusion equation of the form:
\begin{align}
\begin{split}
p(x,y,z,t)=\sum_{i,j,k}^{}A_{ijk} e^{-D\pi^2\Big(\frac{i^2}{x_c^2}+\frac{j^2}{y_c^2}+\frac{k^2}{z_c^2} \Big)t}\text{sin}\Big(\frac{i\pi x}{x_c}\Big)&\text{sin}\Big(\frac{j\pi y}{y_c}\Big)\text{sin}\Big(\frac{k\pi z}{z_c}\Big)
\end{split}
\end{align}
where $x_c$, $y_c$, and $z_c$ are the Cartesian dimensions of the cell chamber, and the mode amplitudes ($A_{ijk}$) are determined by the initial spin distribution ($p(x,y,z,0)$) configured by the pumping beam. If only the fundamental mode ($i,j,k=1$) is excited, then the expected spin decay rate is given by 
\begin{equation}
\label{eq:smFundMode}
    1/\tau_{\text{wall}}=\frac{\pi^2D_0P_0}{P_{N_2}}\frac{1}{\frac{1}{x_c^2}+\frac{1}{y_c^2}+\frac{1}{z_c^2}}
\end{equation}

To model wall collisions in our measurements, it is not necessary to know which higher-order modes are excited, rather we use a phenomenological model where we assume only the fundamental diffusion mode (Eq.~\eqref{eq:smFundMode}) but with a time-dependent buffer-gas pressure $P_{\text{N}_2}(t)\propto \tau_{\text{wall}}(t)$. We model the time-dependence of $P_{\text{N}_2}(t)$ as proportional to the measured instantaneous spin-decay rate shown in Fig.~\ref{fig:smWallColl}(b), where we subdivide this signal into sections of $\Delta t= 35$ $\mu$s and fit a single exponential time constant over each of these time intervals and plot these time-constants over the duration of the spin decay. In this framework, the spin-decay rate at $t \rightarrow \infty$ is given by the fundamental diffusion mode determined by an effective buffer-gas pressure likely to be an underestimate since residual light scattering is lumped into this wall-collision rate.

\subsection{Initial atomic state estimation}
\begin{figure*}[tbh]
\includegraphics[width=0.9\textwidth]{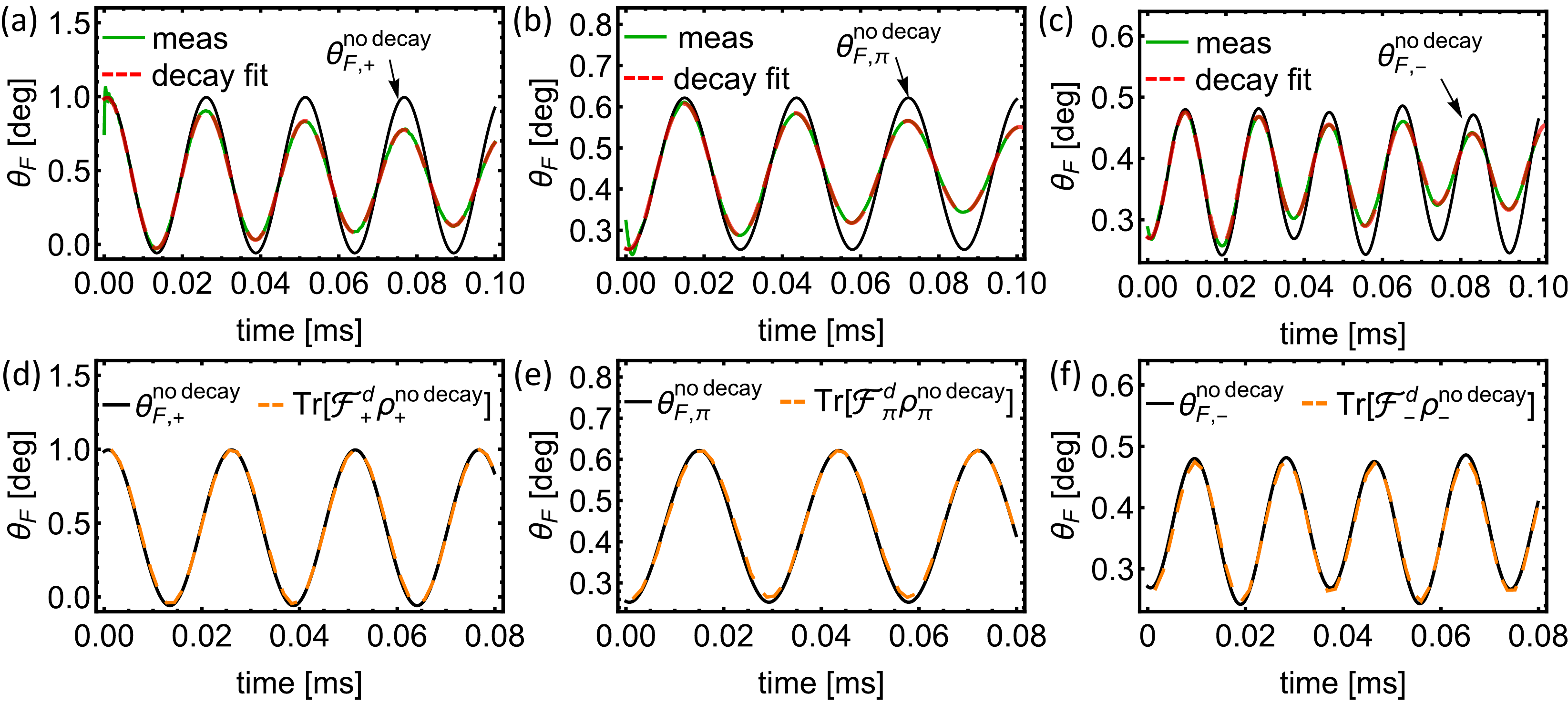}
\caption{Estimate of atomic dynamics during Rabi oscillation without decoherence. (a-c) An exponential decaying sine (red-dashed) is fitted to the measured Faraday rotation signal (green). The decay constants in these fits are set to zero to estimate the Rabi oscillation signal without decoherence $\theta_{F,j}^{\text{no decay}}$ (black). (d-e) To estimate the initial atomic state prior to each Rabi oscillation, we fit the atomic dynamics without decoherence Tr[$\mathcal{F}^d_j \rho_j^{\text{no decay}}$] (orange-dashed) to $\theta_{F,j}^{\text{no decay}}$ (black). }
\label{fig:Figure3supp}
\end{figure*}
The lineshapes of the Rabi oscillations measured in this work are very sensitive to the initial atomic state. There are two initial states to consider: (1) the state after optical pumping $\rho^{\text{OP}}$ that initializes the $\sigma^+$ Rabi oscillation, and (2) the state after adiabatic rapid passage $\rho^{\text{ARP}}$ that initializes the $\sigma^-$ and $\pi$ Rabi oscillations. To simplify these initial states, we assume that atomic coherences are set to zero such that the task at hand is to find the atomic populations $\rho^{\text{OP}}_{ii}$ and $\rho^{\text{ARP}}_{ii}$ 

To incorporate the effect of spin-decay that occurs during the $700$ $\mu$s ARP pulse, we model $\rho^{\text{ARP}}$ in terms of the decayed atomic populations 
\begin{equation}
    \tilde{\rho}_{ii}^{\text{OP}}= (\rho_{ii}^{\text{OP}}-1/8)e^{-s}+1/8
\end{equation}
Here, we set $e^{-s}\approx0.78$ to account for the atomic spin polarization decreasing by 22\%  during the initial $700$ $\mu$s after optical pumping  as measured from the data in Fig.~\ref{fig:smWallColl}. We assume that the primary effect of ARP is to flip the $\ket{2,2}\leftrightarrow \ket{1,1}$ atomic populations with efficiency $\epsilon$, while having no effect on the other atomic populations. We write this quantitatively as
\begin{equation}
\begin{split}
    \rho_{11}^{\text{ARP}}=\tilde{\rho}_{11}^{\text{OP}}+\epsilon(\tilde{\rho}_{44}^{\text{OP}}-\tilde{\rho}_{11}^{\text{OP}})\\
     \rho_{44}^{\text{ARP}}=\tilde{\rho}_{44}^{\text{OP}}-\epsilon(\tilde{\rho}_{44}^{\text{OP}}-\tilde{\rho}_{11}^{\text{OP}})
\end{split}
\end{equation}
 and for all other populations $\rho_{ii}^{\text{ARP}}=\tilde{\rho}_{ii}^{\text{OP}}$. Thus, finding unique solutions for $\rho_{ii}^{\text{OP}}$ and $\epsilon$ fully defines $\rho^{\text{OP}}$ and $\rho^{\text{ARP}}$ with this model.
\begin{figure}[tbh]
\includegraphics[width=0.45\textwidth]{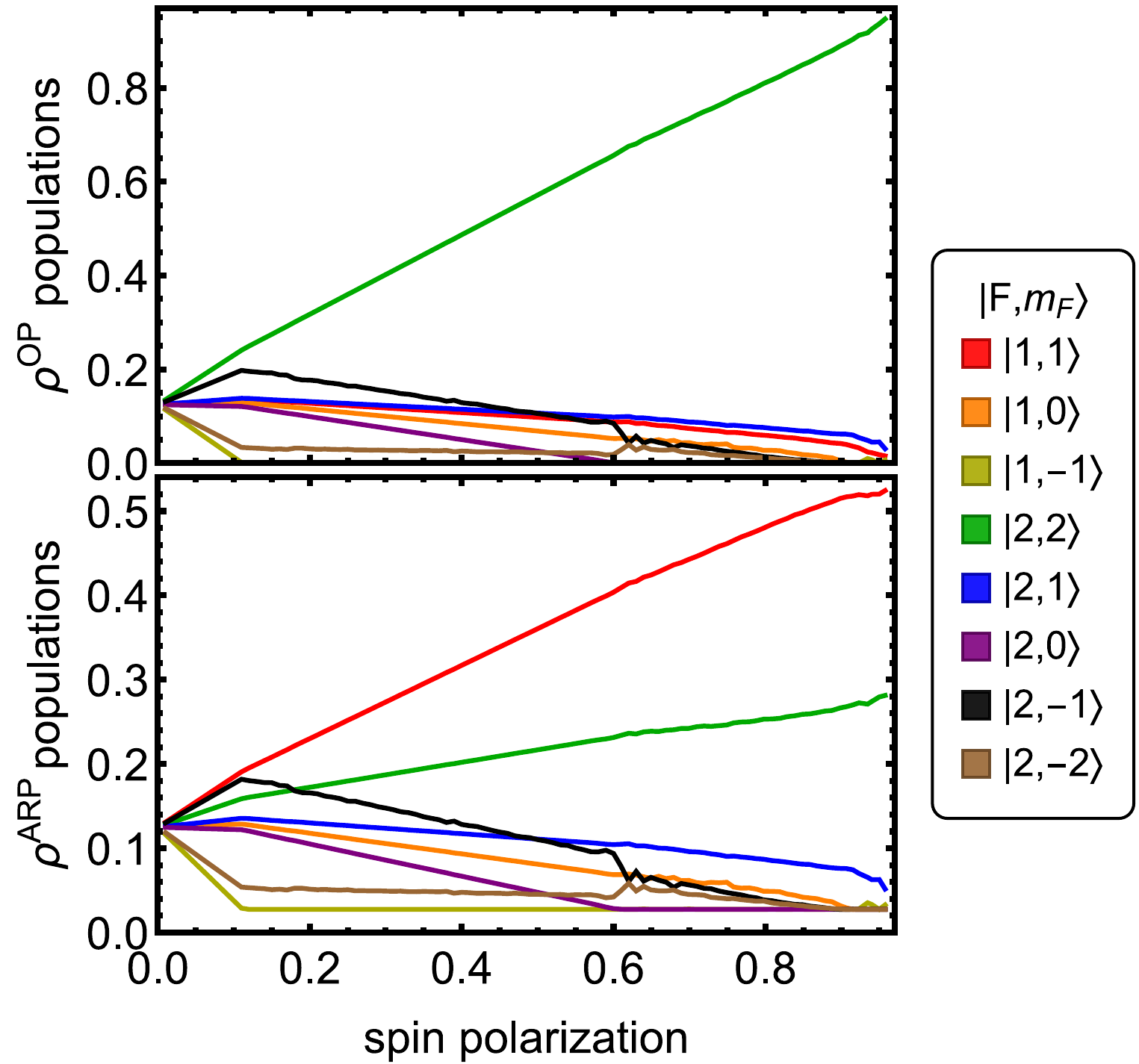}
\caption{Fitted atomic populations subsequent to optical pumping (top) and adiabatic-rapid-passage (bottom) as a function of the electron spin polarization (p) produced at the end of optical pumping}
\label{fig:Figure4supp}
\end{figure}

Next, we show that we can fit $\rho_{ii}^{\text{op}}$ from our Rabi data. To do this, we first extract an estimate for the Rabi oscillations without decoherence as shown in Fig.~\ref{fig:Figure3supp}. Here, we fit an exponential decay model $\theta_F(t)=A_0+A_1 e^{-t/T_1}+A_2e^{-t/T_2}\text{cos}(2\pi f t)$ to the initial 100 $\mu$s of each Rabi oscillation, and use the $\theta^{\text{no decay}}_F(t)=A_0+A_1 +A_2\text{cos}(2\pi f t)$ as an estimate for the Rabi oscillation without decoherence.
 
From fitting to the Rabi data in Fig~\ref{fig:smRabiACStark}, we apriori know the full time-independent Hamiltonian $\tilde{H}$ for each Rabi oscillation. We diagonalize $\tilde{H}\rightarrow \tilde{H}_d$ and transform the initial atomic state $\rho_0=\rho^{\text{OP}}$ (or $\rho^{\text{ARP}}$) into this diagonalized basis $\rho_0 \rightarrow \rho_{0,d}$. Then the atomic dynamics without decoherence is given by \begin{equation}
 \rho^{\text{no decay}}(t)=e^{-i\tilde{H}_dt/\hbar}\rho_{0,d}e^{i\tilde{H}_dt/\hbar} 
\end{equation}
By writing $\rho_{0,d}$ as a function of the unknown atomic populations $\rho_{jj}^{\text{OP}}$, we can fit $a_0$Tr$[\mathcal{F}_d\rho^{\text{no decay}}(t)]+b_0$ to $\theta^{\text{no decay}}(t)$ for each Rabi oscillation to extract $\rho_{jj}^{\text{OP}}$ and $\epsilon$. Here the Faraday rotation offset $b_0$ is known from the polarimeter reading prior to optical pumping, but the scaling variable $a_0$ is an additional fitting parameter. The cost function that we minimize is
\begin{equation}
\label{eq:costFun}
\begin{split}
    \sum_{r=\pm,-\pi}\sum_{k=1}^{50}& w_r\Big[b_0+a_0\text{Tr}[\mathcal{F}^d_r\rho_r^{\text{no decay}}(t_k)]-\theta_r^{\text{no decay}}(t_k)\Big]^2
    \end{split}
\end{equation}
that samples 50 sequential time steps seperated by $t_{k+1}-t_k=1.6$ $\mu$s. Here $w_+=1$, $w_{\pi}=2=A_{2,+}/A_{2,\pi}$, and $w_-=4=A_{2,+}/A_{2,-}$ are weights chosen based on the fitting amplitude $A_2$ of each Rabi oscillation in $\theta^{\text{no decay}}(t)$. The results of this fitting is shown in Fig.~\ref{fig:Figure3supp}. We find unique solutions for $\rho_{jj}^{\text{op}}$ if we also fix the atomic spin polarization $p$ shown in Fig.~\ref{fig:Figure4supp}. 

\subsection{Extracting the vapor temperature, buffer gas pressure, and electron spin polarization from Rabi oscillations}

\begin{figure*}[tbh]
\includegraphics[width=0.99\textwidth]{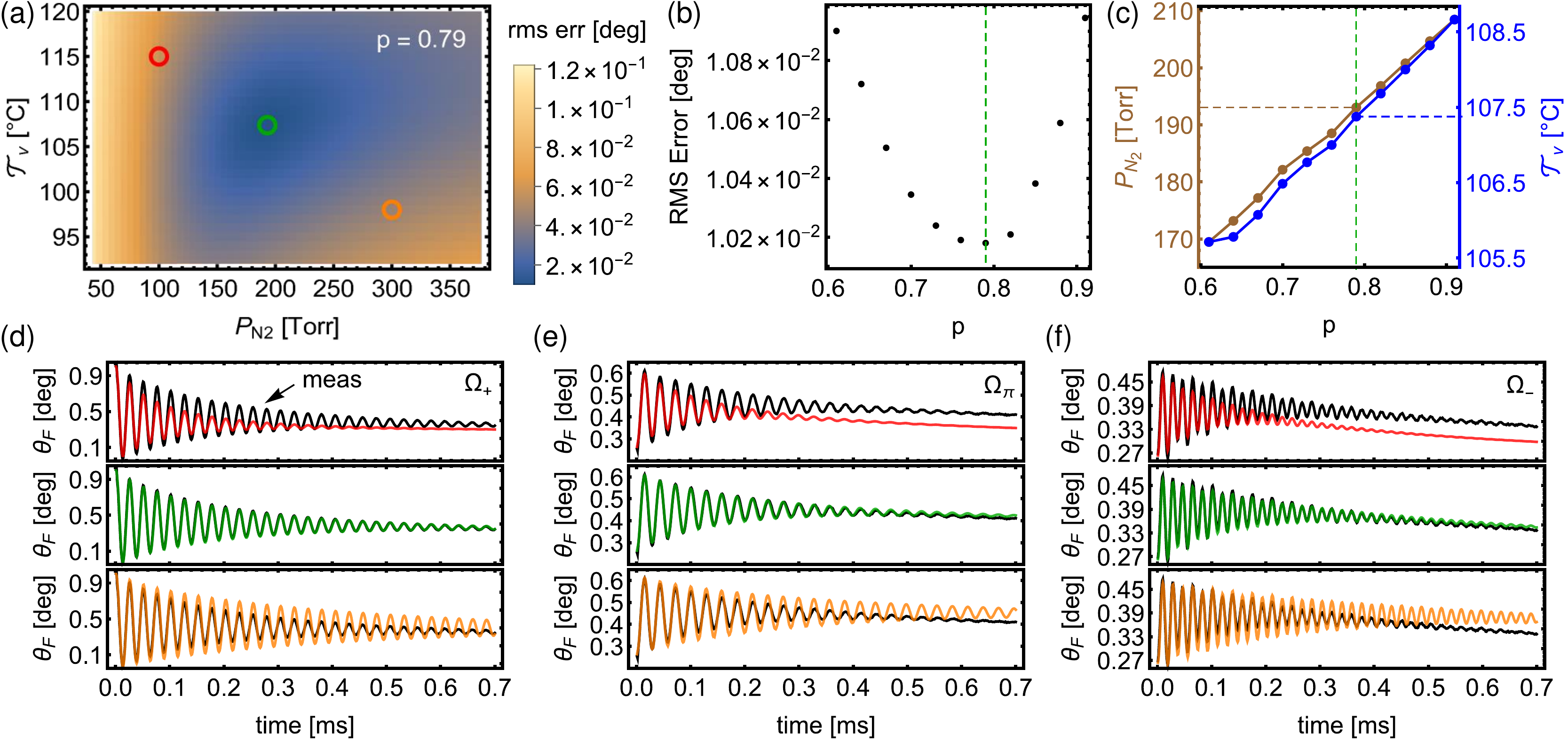}
\caption{Fitting vapor cell parameters (a) The RMS error $e_{\theta_F}$ between the simulated and measured Rabi oscillations as a function of vapor temperature $\mathcal{T}_v$ and buffer gas pressure P$_{\text{N}_2}$ for an electron spin polarization of p = 0.79. Circles denote optimal (green) and non-optimal (red and orange) vapor cell parameters within the theoretical model (b) the RMS error $e_{\theta_F}$ at optimal $\mathcal{T}_v$ and P$_{\text{N}_2}$ for different spin polarization p. (c) Optimal $\mathcal{T}_v$ and P$_{\text{N}_2}$ values as a function of the electron spin polarization. (d-f) Comparison of measured (black) and simulated Rabi oscillations for different vapor parameters given by the circles in (a). Optimal values are the green center curves.}
\label{fig:Figure5supp}
\end{figure*}

While we now have an estimate for the initial atomic state prior to each Rabi oscillation, we still have unknown parameters in our model such as the vapor temperature $\mathcal{T}_v$, the buffer gas pressure P$_{\text{N}_2}$, and the electron spin polarization produced by optical pumping $p$. We restate the theoretical model in the main text that we are interested in solving
\begin{align}
\label{eq:timeEvoSupp}
\begin{split}
    \frac{d\rho}{dt}=&\frac{[\tilde{H},\rho]}{i \hbar}+\Gamma_{\text{se}}\big(\tilde{\phi}(1+4\langle \tilde{\mathbf{S}} \rangle\cdot \tilde{\textbf{S}})-\rho\big)+\Gamma_{\text{sd}}\big(\tilde{\phi}-\rho\big)-\frac{\eta_I^2 [I]}{8}\Gamma_{\text{C}} \rho^{(\text{m})}+  \Gamma_D(t)(\rho^e-\rho)
\end{split}
\end{align}
where $\tilde{\mathcal{O}}$ denote operators made time-independent with the rotating-wave approximation and $\Gamma_D(t)$ accounts for time-dependent spin decay that we measure independently in Fig.~\ref{fig:smWallColl} that we model as arising from spin diffusion discussed in a previous section. The parameters $\{\mathcal{T}_v,\text{P}_{\text{N}_2},p\}$ affect both the Rabi coherence through the collision rates and the overall population dynamics. To find the optimal values for these parameters within our model, we minimize the RMS error
\begin{equation}
e_{\theta_F}=\sqrt{\frac{\sum_{r=\pm,\pi}\sum_{j=1}^N(\theta_{F,r}^{\text{sim}}(t_j)-\theta_{F,r}(t_j))^2}{3N}}\end{equation}
by utilizing the lineshapes of all three hyperfine transitions at sampled times $t_j$ where $t_{j+1}-t_j=1$ $\mu$s. Here $\theta_F(t)$ is the measured Faraday rotation and $\theta^{\text{sim}}_F(t)$  is the full-simulated Rabi oscillation Faraday signal. In $\theta^{\text{sim}}_F(t)$ we use the same scaling factor $a_0$ fitted during the extraction of initial atomic populations (see Eq.~\eqref{eq:costFun}). We minimize $e_{\theta_F}$ by evaluating $e_{\theta_F}$ over a 2D scan of $\mathcal{T}_v$ and $\text{P}_{\text{N}_2}$ at a given electron spin polarization $p$ (Fig.~\ref{fig:Figure5supp}a). We use piecewise polynomial interpolation of order 3 to estimate the global minimum appearing in these 2D scans shown as a green circle in Fig.~\ref{fig:Figure5supp}a. We plot the global minimum of $e_{\theta_F}$ at different $p$ in Fig.~\ref{fig:Figure5supp}b and the corresponding values of $\mathcal{T}_v$ and $P_{\text{N}_2}$ in Fig.~\ref{fig:Figure5supp}(b-c), where the vertical green-dashed line indicates the optimal values. An overlay of the simulated with the measured Rabi oscillations for different vapor parameters (colored circles in Fig.~\ref{fig:Figure5supp}a) at an electron spin polarization $p=0.79$ is shown in Fig.~\ref{fig:Figure5supp}(d-f) to visually illustrate how unique values of $\{\mathcal{T}_v,\text{P}_{\text{N}_2},p\}$ fit our data.

\section{Independent measurements of buffer-gas pressure}
Buffer-gas collisions cause frequency shifts and broadening in both microwave and optical transitions.  Here we measure the buffer gas pressure from the microwave frequency shift $\delta_{\text{hfs}}$, extracted from Rabi measurements in this work. As another reference, we also measure the optical broadening and frequency shift in a separate setup to extract two additional measurements of the $\text{N}_2$ buffer gas pressure.
\subsection{Ground state frequency shift from Rabi measurements}
Perturbations to the alkali electron clouds cause a variation of the hyperfine coupling constant $\delta A_{\text{hfs}} \mathbf{S}\cdot \mathbf{I}$, which results in a frequency shift to the hyperfine splitting~\cite{vanier1989quantum}
\begin{equation}\delta \nu_{\text{hfs}}=\text{P}_0\big[\beta_0+\delta_0(\mathcal{T}_v-\mathcal{T}_{v,0})+O(\mathcal{T}_v^2) \big]
\end{equation}
where $\text{P}_0=n_{\text{N}_2}k_B \mathcal{T}_{v,0}$ is the buffer-gas pressure at reference temperature $\mathcal{T}_{v,0}$. For $^{87}$Rb with $\text{N}_2$ buffer gas measured at reference temperature $\mathcal{T}_{v,0}=60\degree$C~\cite{missout1975pressure}
\begin{align}
\begin{split}
\beta_0/\nu_{0}=81.9\pm1.2 \times 10^{-9} \text{ Torr}^{-1}\\
\delta_0/\nu_{0}=79\pm 2 \times 10^{-12} \text{ }C^{-1}\text{Torr}^{-1}
\end{split}
\end{align}

The actual buffer gas pressure for temperature $\mathcal{T}_v$ is $\text{P}_{N_2}=n_{\text{N}_2}k_B \mathcal{T}_v=\text{P}_0\frac{\mathcal{T}_v}{\mathcal{T}_{v,0}}$. Thus,
\begin{equation}
\text{P}_{N_2}=\frac{\mathcal{T}_v}{\mathcal{T}_{v,0}}\frac{\delta \nu_{\text{hfs}}}{\beta_0+\delta_0(\mathcal{T}_v-\mathcal{T}_{v,0})}
\end{equation}
A fitted frequency shift $\delta \nu_{\text{hfs}}=87.22$ kHz (Fig.~\ref{fig:smRabiACStark}) implies $\text{P}_{\text{N}_2}=168$ Torr at $\mathcal{T}_v=373.15$K.

\subsection{D1 optical frequency shift and broadening}
\begin{figure}[tbh]
\includegraphics[width=0.45\textwidth]{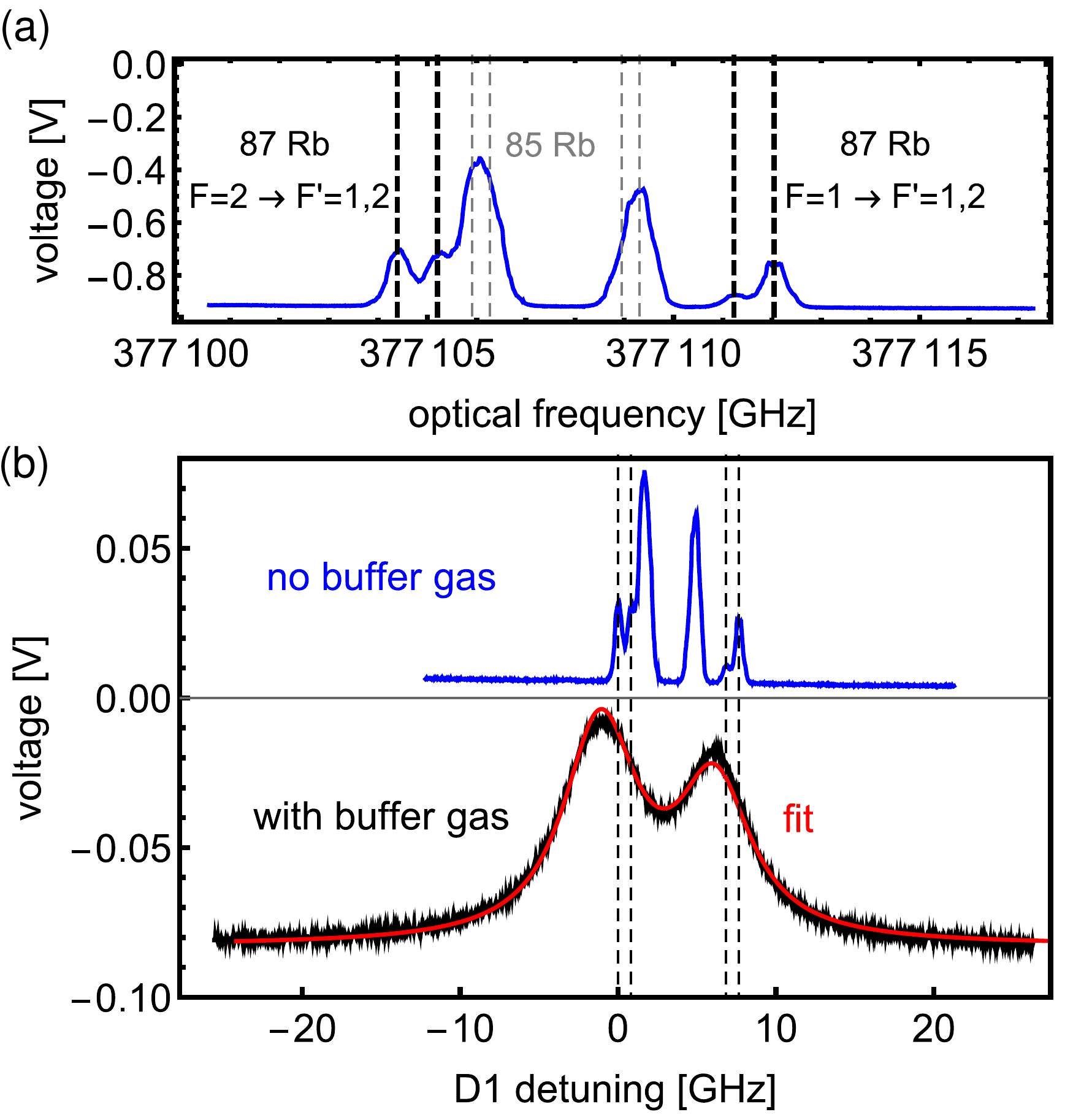}
\caption{ (a) Transmission spectrum with a vapor cell containing no buffer gas used to calibrate the optical frequency~\cite{steckrubidium}. Dashed lines mark $^{87}$Rb (black) and $^{85}$Rb (gray) D1 transitions. (b) Transmission through a microfabricated vapor cell with $\text{N}_2$ buffer gas (black) that is measured simultaneously with (a).}
\label{fig:smPressBroad}
\end{figure}
The buffer gas pressure can also be extracted from the linewidths and frequency shifts of optical transitions. We use a 795 nm VCSEL to measure the D1 optical spectrum through a heated microfabricated cell with $\text{N}_2$ buffer gas used in this work, and a glass cell that contains no buffer gas but serves as an optical frequency reference. We attenuate the VCSEL power through the microfabricated cell to $6.8$ $\mu$W (beam waist $\approx$ 2 mm) to minimize power broadening. We verified that decreasing the beam power further to $0.7$ $\mu$W had little effect on the optical broadening.

A linear triangle frequency ramp sweeps the VCSEL optical frequency by changing its temperature. From the transmission through the glass cell, the $^{87}$Rb and $^{85}$Rb transitions are resolved to calibrate the optical frequency, as shown in Fig~\ref{fig:smPressBroad}, by fitting a Lorentizian lineshape 
\begin{align}
\begin{split}
    L(f,f_{0,F,F'},A_{F,F'},\gamma)=
    A_{F,F'}&\frac{\gamma/2}{(f-f_{0,F,F'})^2+(\gamma/2)^2}+\text{off}
\end{split}
\end{align}
to each of the four $^{87}$Rb resonances, where $f_{0,F,F'}$, $A_{F,F'}$, and $\gamma$ are fitting parameters for the resonant frequency, amplitude, and a common width for each resonance. We also include an offset due to bias on the photodiode. Here $F,F'$ denotes the specific fitted parameter for the $F\rightarrow F'$ transition.

To fit the pressure-broadened D1 spectrum we use the following fitting function
\begin{equation}
    A_0\sum_{F,F'=1}^{F,F'=2} L(f,f_{0,F,F'}+f_{N_2},A_{F,F'},\gamma_{N_2})+B_0f +\text{off}
\end{equation}
where $A_{F,F'}$ and $f_{0,F,F'}$ are the fitted amplitudes and frequencies from the spectrum with no buffer gas. The only fitting parameters here are $A_0$, $B_0$, off, $f_{\text{N}_2}= 1.08$ GHz and $\gamma_{\text{N}_2}= 5.56$ GHz.

The broadening and shift are given by $f_{\text{N}_2}=\alpha n_{\text{N}_2}$ and $\gamma_{\text{N}_2}=\alpha^{\prime}n_{\text{N}_2}$, where $\alpha=17.8\pm0.3$ GHz/amg, $\alpha^{\prime}=-8.25\pm0.15$ GHz/amg, and $n_{\text{N}_2}=\frac{\text{P}_{\text{N}_2}}{44.615 N_A k_B T }$ is the buffer gas density in amg units using Avogadro's number $N_A$~\cite{romalis1997pressure}. From the fitted values for $f_{\text{N}_2}$ and $\gamma_{\text{N}_2}$, we obtain $\text{P}_{\text{N}_2}=182\pm 4$ Torr and $\text{P}_{\text{N}_2}=332\pm 4$ Torr respectively.

We suspect that part of the discrepancy between these buffer-gas pressures is due to the contamination of hydrocarbons during the in-house manufacturing of the vapor cell. In this case, the extracted N$_2$ buffer-gas pressure could vary between independent measurements since hydrocarbon molecules do not have the same frequency shifts and broadening coefficients ($\alpha$,$\alpha^{\prime}$) as $\text{N}_2$. For example, the D1 optical broadening coefficient for CH$_4$ is nearly twice as large as the broadening coefficient for N$_2$, while the D1 frequency-shift coefficients are nearly equal~\cite{rotondaro1997collisional}. The biggest impact on Rabi coherence from unknown hydrocarbon contamination would be a perturbation to the expected wall-collision rate. Since, we independently measure spin decay (Fig.~\ref{fig:smWallColl}) and incorporate this information as a phenomenological wall-collision rate, we account for the exact spin decay in our experiment without appealing to the specific collisional mechanism. Thus, hydrocarbon contamination only affects the absolute pressure shift we extract, and not the main result of the paper --- a full understanding of Rabi coherence from spin-exchange collisions. 

\begin{table*}
\caption{\label{elecSpin}%
Electron spin matrices in the $\ket{m_s,m_I}$ and  $\ket{F,m_F}$ bases for $^{87}$Rb.}
\begin{ruledtabular}
\begin{tabular}{c c c}
\textrm{Spin} &
\textrm{$\ket{m_s,m_I}$} &
\textrm{$\ket{F,m_F}$}\\[1mm]
\textrm{Operator} &
\textrm{\tiny $\ket{-\frac{1}{2},\frac{1}{2}}$ $\ket{-\frac{1}{2},-\frac{1}{2}}$ $\ket{-\frac{1}{2},-\frac{3}{2}}$ $\ket{\frac{1}{2},\frac{3}{2}}$ $\ket{\frac{1}{2},\frac{1}{2}}$ $\ket{\frac{1}{2},-\frac{1}{2}}$ $\ket{\frac{1}{2},-\frac{3}{2}}$ $\ket{-\frac{1}{2},\frac{3}{2}}$} &
\textrm{\scriptsize$ \ket{1,1}$ $\ket{1,0}$ $\ket{1,-1}$ $\ket{2,2}$ $\ket{2,1}$ $\ket{2,0}$ $\ket{2,-1}$ $\ket{2,-2}$ }\\[1mm]
\colrule
\\ [-.3em]
 $S_x$&$\left[\begin{array}{*8{C{2em}}}
   0&0&0&0&\frac{1}{2}&0&0&0\\[2.2mm]
   0&0&0&0&0&\frac{1}{2}&0&0\\[2.2mm]
   0&0&0&0&0&0&\frac{1}{2}&0\\[2.2mm]
  0&0&0&0&0&0&0&\frac{1}{2}\\[2.2mm]
   \frac{1}{2}&0&0&0&0&0&0&0\\[2.2mm]
   0&\frac{1}{2}&0&0&0&0&0&0\\[2.2mm]
   0&0&\frac{1}{2}&0&0&0&0&0\\[2.2mm]
   0&0&0&\frac{1}{2}&0&0&0&0
  \end{array}\right]$& 
$\left[\begin{array}{*8{C{2em}}}
   0&-\frac{1}{4\sqrt{2}}&0&-\frac{\sqrt{3}}{4}&0&\frac{1}{4\sqrt{2}}&0&0\\[1mm]
   -\frac{1}{4\sqrt{2}}&0&-\frac{1}{4\sqrt{2}}&0&-\frac{\sqrt{\frac{3}{2}}}{4}&0&\frac{\sqrt{\frac{3}{2}}}{4}&0\\[1mm]
   0&-\frac{1}{4\sqrt{2}}&0&0&0&-\frac{1}{4\sqrt{2}}&0&\frac{\sqrt{3}}{4}\\[1mm]
  -\frac{\sqrt{3}}{4}&0&0&0&\frac{1}{4}&0&0&0\\[1mm]
   0&-\frac{\sqrt{\frac{3}{2}}}{4}&0&\frac{1}{4}&0&\frac{\sqrt{\frac{3}{2}}}{4}&0&0\\[1mm]
   \frac{1}{4\sqrt{2}}&0& -\frac{1}{4\sqrt{2}}&0&\frac{\sqrt{\frac{3}{2}}}{4}&0&\frac{\sqrt{\frac{3}{2}}}{4}&0\\[1mm]
   0&\frac{\sqrt{\frac{3}{2}}}{4}&0&0&0&\frac{\sqrt{\frac{3}{2}}}{4}&0&\frac{1}{4}\\[1mm]
   0&0&\frac{\sqrt{3}}{4}&0&0&0&\frac{1}{4}&0
  \end{array}\right]$\\
\\ [2mm]
$S_y$&$\left[\begin{array}{*8{C{2em}}}
   0&0&0&0&\frac{i}{2}&0&0&0\\[2.2mm]
   0&0&0&0&0&\frac{i}{2}&0&0\\[2.2mm]
   0&0&0&0&0&0&\frac{i}{2}&0\\[2.2mm]
  0&0&0&0&0&0&0&-\frac{i}{2}\\[2.2mm]
   -\frac{i}{2}&0&0&0&0&0&0&0\\[2.2mm]
   0&-\frac{i}{2}&0&0&0&0&0&0\\[2.2mm]
   0&0&-\frac{i}{2}&0&0&0&0&0\\[2.2mm]
   0&0&0&\frac{i}{2}&0&0&0&0
  \end{array}\right]$ &$\left[\begin{array}{*8{C{2em}}}
   0&\frac{i}{4\sqrt{2}}&0&-\frac{i\sqrt{3}}{4}&0&-\frac{i}{4\sqrt{2}}&0&0\\[1mm]
   -\frac{i}{4\sqrt{2}}&0&\frac{i}{4\sqrt{2}}&0&-\frac{i\sqrt{\frac{3}{2}}}{4}&0&-\frac{i\sqrt{\frac{3}{2}}}{4}&0\\[1mm]
   0&-\frac{i}{4\sqrt{2}}&0&0&0&-\frac{i}{4\sqrt{2}}&0&-\frac{i\sqrt{3}}{4}\\[1mm]
  \frac{i\sqrt{3}}{4}&0&0&0&-\frac{i}{4}&0&0&0\\[1mm]
   0&\frac{i\sqrt{\frac{3}{2}}}{4}&0&\frac{i}{4}&0&-\frac{i\sqrt{\frac{3}{2}}}{4}&0&0\\[1mm]
   \frac{i}{4\sqrt{2}}&0& \frac{i}{4\sqrt{2}}&0&\frac{i\sqrt{\frac{3}{2}}}{4}&0&-\frac{i\sqrt{\frac{3}{2}}}{4}&0\\[1mm]
   0&\frac{i\sqrt{\frac{3}{2}}}{4}&0&0&0&\frac{i\sqrt{\frac{3}{2}}}{4}&0&-\frac{i}{4}\\[1mm]
   0&0&\frac{i\sqrt{3}}{4}&0&0&0&\frac{i}{4}&0
  \end{array}\right]$\\
\\ [2mm]
$S_z$&$\left[\begin{array}{*8{C{2em}}}
   -\frac{1}{2}&0&0&0&0&0&0&0\\[2mm]
   0&-\frac{1}{2}&0&0&0&0&0&0\\[2mm]
   0&0&-\frac{1}{2}&0&0&0&0&0\\[2mm]
  0&0&0&\frac{1}{2}&&0&0&0\\[2mm]
   0&0&0&0&\frac{1}{2}&0&0&0\\[2mm]
   0&0&0&0&0&\frac{1}{2}&0&0\\[2mm]
   0&0&0&0&0&0&\frac{1}{2}&0\\[2mm]
   0&0&0&0&0&0&0&-\frac{1}{2}\\[2mm]
  \end{array}\right]$ &$\left[\begin{array}{*8{C{2em}}}
   -\frac{1}{4}&0&0&0&\frac{\sqrt{3}}{4}&0&0&0\\[2mm]
   0&0&0&0&0&\frac{1}{2}&0&0\\[2mm]
   0&0&\frac{1}{4}&0&0&0&\frac{\sqrt{3}}{4}&0\\[2mm]
  0&0&0&\frac{1}{2}&0&0&0&0\\[2mm]
   \frac{\sqrt{3}}{4}&0&0&0&\frac{1}{4}&0&0&0\\[2mm]
   0&\frac{1}{2}&0&0&0&0&0&0\\[2mm]
   0&0&\frac{\sqrt{3}}{4}&0&0&0&-\frac{1}{4}&0\\[2mm]
   0&0&0&0&0&0&0&-\frac{1}{2}
  \end{array}\right]$\\[1mm]
\end{tabular}
\end{ruledtabular}
\end{table*}
\clearpage
\begin{table*}
\caption{\label{nucSpin}%
Nuclear spin matrices in the $\ket{m_s,m_I}$ and  $\ket{F,m_F}$ bases for $^{87}$Rb.}
\begin{ruledtabular}
\begin{tabular}{c c c}
\textrm{Spin} &
\textrm{$\ket{m_s,m_I}$} &
\textrm{$\ket{F,m_F}$}\\[1mm]
\textrm{Operator} &
\textrm{\tiny $\ket{-\frac{1}{2},\frac{1}{2}}$ $\ket{-\frac{1}{2},-\frac{1}{2}}$ $\ket{-\frac{1}{2},-\frac{3}{2}}$ $\ket{\frac{1}{2},\frac{3}{2}}$ $\ket{\frac{1}{2},\frac{1}{2}}$ $\ket{\frac{1}{2},-\frac{1}{2}}$ $\ket{\frac{1}{2},-\frac{3}{2}}$ $\ket{-\frac{1}{2},\frac{3}{2}}$} &
\textrm{\scriptsize$ \ket{1,1}$ $\ket{1,0}$ $\ket{1,-1}$ $\ket{2,2}$ $\ket{2,1}$ $\ket{2,0}$ $\ket{2,-1}$ $\ket{2,-2}$ }\\[1mm]
\colrule
\\ [-.3em]
$I_x$&$\left[\begin{array}{*8{C{2em}}}
   0&1&0&0&0&0&0&\frac{\sqrt{3}}{2}\\[1.8mm]
   1&0&\frac{\sqrt{3}}{2}&0&0&0&0&0\\[1.8mm]
   0&\frac{\sqrt{3}}{2}&0&0&0&0&0&0\\[1.8mm]
  0&0&0&0&\frac{\sqrt{3}}{2}&0&0&0\\[1.8mm]
   0&0&0&\frac{\sqrt{3}}{2}&0&1&0&0\\[1.8mm]
   0&0&0&0&1&0&\frac{\sqrt{3}}{2}&0\\[1.8mm]
   0&0&0&0&0&\frac{\sqrt{3}}{2}&0&0\\[1.8mm]
   \frac{\sqrt{3}}{2}&0&0&0&0&0&1&0
   \end{array}\right]$ &$\left[\begin{array}{*8{C{2em}}}
   0&\frac{5}{4\sqrt{2}}&0&\frac{\sqrt{3}}{4}&0&-\frac{1}{4\sqrt{2}}&0&0\\[1.2mm]
   \frac{5}{4\sqrt{2}}&0&\frac{5}{4\sqrt{2}}&0&\frac{\sqrt{\frac{3}{2}}}{4}&0&-\frac{\sqrt{\frac{3}{2}}}{4}&0\\[1.2mm]
   0&\frac{5}{4\sqrt{2}}&0&0&0&\frac{1}{4\sqrt{2}}&0&-\frac{\sqrt{3}}{4}\\[1.2mm]
  \frac{\sqrt{3}}{4}&0&0&0&\frac{3}{4}&0&0&0\\[1.5mm]
   0&\frac{\sqrt{\frac{3}{2}}}{4}&0&\frac{3}{4}&0&\frac{3\sqrt{\frac{3}{2}}}{4}&0&0\\[1.2mm]
   -\frac{1}{4\sqrt{2}}&0&\frac{1}{4\sqrt{2}}&0&\frac{3\sqrt{\frac{3}{2}}}{4}&0&\frac{3\sqrt{\frac{3}{2}}}{4}&0\\[1.2mm]
   0&-\frac{\sqrt{\frac{3}{2}}}{4}&0&0&0&\frac{3\sqrt{\frac{3}{2}}}{4}&0&\frac{3}{4}\\[1.2mm]
   0&0&-\frac{\sqrt{3}}{4}&0&0&0&\frac{3}{4}&0
   \end{array}\right]$\\
  \\[2mm]
$I_y$&$\left[\begin{array}{*8{C{2em}}}
   0&-i&0&0&0&0&0&\frac{i\sqrt{3}}{2}\\[1.8mm]
   i&0&-\frac{i\sqrt{3}}{2}&0&0&0&0&0\\[1.8mm]
   0&\frac{i\sqrt{3}}{2}&0&0&0&0&0&0\\[1.8mm]
  0&0&0&0&-\frac{i\sqrt{3}}{2}&0&0&0\\[1.8mm]
   0&0&0&\frac{i\sqrt{3}}{2}&0&-i&0&0\\[1.8mm]
   0&0&0&0&i&0&-\frac{i\sqrt{3}}{2}&0\\[1.8mm]
   0&0&0&0&0&\frac{i\sqrt{3}}{2}&0&0\\[1.8mm]
   -\frac{i\sqrt{3}}{2}&0&0&0&0&0&0&0
   \end{array}\right]$ &$\left[\begin{array}{*8{C{2em}}}
   0&-\frac{i5}{4\sqrt{2}}&0&\frac{i\sqrt{3}}{4}&0&-\frac{i}{4\sqrt{2}}&0&0\\[1.2mm]
   \frac{i5}{4\sqrt{2}}&0&-\frac{i5}{4\sqrt{2}}&0&\frac{i\sqrt{\frac{3}{2}}}{4}&0&\frac{i\sqrt{\frac{3}{2}}}{4}&0\\[1.2mm]
   0&\frac{i5}{4\sqrt{2}}&0&0&0&\frac{i}{4\sqrt{2}}&0&\frac{i\sqrt{3}}{4}\\[1.2mm]
  -\frac{i\sqrt{3}}{4}&0&0&0&-\frac{i3}{4}&0&0&0\\[1.2mm]
   0&-\frac{i\sqrt{\frac{3}{2}}}{4}&0&\frac{i3}{4}&0&-\frac{i3\sqrt{\frac{3}{2}}}{4}&0&0\\[1.2mm]
   -\frac{i}{4\sqrt{2}}&0&-\frac{i}{4\sqrt{2}}&0&\frac{i3\sqrt{\frac{3}{2}}}{4}&0&-\frac{i3\sqrt{\frac{3}{2}}}{4}&0\\[1.2mm]
   0&-\frac{i\sqrt{\frac{3}{2}}}{4}&0&0&0&\frac{i3\sqrt{\frac{3}{2}}}{4}&0&-\frac{i3}{4}\\[1.2mm]
   0&0&-\frac{i\sqrt{3}}{4}&0&0&0&\frac{i3}{4}&0
   \end{array}\right]$\\
  \\[2mm]
$I_z$&$\left[\begin{array}{*8{C{2em}}}
   \frac{1}{2}&0&0&0&0&0&0&0\\[2mm]
   0&-\frac{1}{2}&0&0&0&0&0&0\\[2mm]
   0&0&-\frac{3}{2}&&0&0&0&0\\[2mm]
  0&0&0&\frac{3}{2}&0&0&0&0\\[2mm]
   0&0&0&0&\frac{1}{2}&0&0&0\\[2mm]
   0&0&0&0&0&-\frac{1}{2}&0&0\\[2mm]
   0&0&0&0&0&0&-\frac{3}{2}&0\\[2mm]
   0&0&0&0&0&0&0&\frac{3}{2}\\[2mm]
   \end{array}\right]$ &$\left[\begin{array}{*8{C{2em}}}
   \frac{5}{4}&0&0&0&-\frac{\sqrt{3}}{4}&0&0&0\\[2mm]
   0&0&0&0&0&-\frac{1}{2}&0&0\\[2mm]
   0&0&-\frac{5}{4}&0&0&0&-\frac{\sqrt{3}}{4}&0\\[2mm]
  0&0&0&\frac{3}{2}&0&0&0&0\\[2mm]
   -\frac{\sqrt{3}}{4}&0&0&0&\frac{3}{4}&0&0&0\\[2mm]
   0&-\frac{1}{2}&0&0&0&0&0&0\\[2mm]
   0&0&-\frac{\sqrt{3}}{4}&0&0&0&-\frac{3}{4}&0\\[2mm]
   0&0&0&0&0&0&0&-\frac{3}{2}
   \end{array}\right]$\\[1mm]
\end{tabular}
\end{ruledtabular}
\end{table*}


%